\documentclass[aps,prx,superscriptaddress,showpacs,reprint]{revtex4-1}
\usepackage{graphicx}
\usepackage{mathrsfs}
\usepackage{amsmath}
\usepackage{subfigure}
\usepackage{bm}
\usepackage{verbatim}
\usepackage[colorlinks=true,letterpaper=true, pdfstartview=FitV,urlcolor=blue,linkcolor=orange,citecolor=cyan]{hyperref}
\usepackage{color}
\usepackage{xcolor}
\usepackage{hyperref}
\usepackage{amssymb}
\usepackage{bbm}
\usepackage{siunitx}
\usepackage{txfonts}
\newcommand{\mH}{\mathcal{H}}
\newcommand{\bq}{\bm{q}}
\newcommand{\bk}{\bm{k}}

\newcommand{\abs}[1]{\lvert #1\rvert}

\newcommand{\mI}{\mathcal{I}}
\newcommand{\mD}{\mathcal{D}}
\newcommand{\average}[1]{\langle#1\rangle}
\newcommand{\schrodinger}{Schr\"{o}dinger equation}
\usepackage{mathtools}

\DeclarePairedDelimiterX\braket[2]{\langle}{\rangle}{#1 \delimsize\vert #2}
\DeclarePairedDelimiterX\inner[2]{\langle}{\rangle}{#1,#2}
\newcommand{\AAM}{Aubry-Andre}

\newcommand{\iLM}{incommensurate lattice model}

\def\be{\begin{equation}}
\def\ee{\end{equation}}
\def\bea{\begin{eqnarray}}
\def\eea{\end{eqnarray}}
\def\nn{\nonumber}

\begin{document}
\title{Mobility Edges in 1D Bichromatic Incommensurate Potentials} 

\author{Xiao Li}
\affiliation{Condensed Matter Theory Center and Joint Quantum Institute, University of Maryland, College Park, Maryland 20742-4111, USA}
\author{Xiaopeng Li}
\affiliation{Condensed Matter Theory Center and Joint Quantum Institute, University of Maryland, College Park, Maryland 20742-4111, USA}
\affiliation{State Key Laboratory of Surface Physics, Institute of Nanoelectronics and Quantum Computing, and Department of Physics, Fudan University, Shanghai 200433, China}
\affiliation{Collaborative Innovation Center of Advanced Microstructures, Nanjing 210093, China}

\author{S. Das Sarma}
\affiliation{Condensed Matter Theory Center and Joint Quantum Institute, University of Maryland, College Park, Maryland 20742-4111, USA}

\date{\today}

\begin{abstract}
We theoretically study a one-dimensional (1D) mutually incommensurate bichromatic lattice system which has been implemented in ultracold atoms to study quantum localization. 
It has been universally believed that the tight-binding version of this bichromatic incommensurate system is represented by the well-known Aubry-Andre model {capturing all the essential localization physics in the experimental cold atom optical lattice system}. Here we establish that this belief is incorrect and that the Aubry-Andre model description, which applies only in the extreme tight-binding limit of very deep primary lattice potential, generically breaks down near the localization transition due to the unavoidable appearance of single-particle mobility edges (SPME). 
In fact, we show that the 1D bichromatic incommensurate potential system manifests generic mobility edges which disappear in the tight-binding limit, leading to the well-studied Aubry-Andre physics. 
We carry out an extensive study of the localization properties of the 1D incommensurate optical lattice without making any tight-binding approximation. 
We find that, for the full lattice system, an intermediate phase between completely localized and completely delocalized regions appears due to the existence of the SPME, making the system qualitatively distinct from the Aubry-Andre prediction. 
Using the Wegner flow approach, we show that the SPME in the real lattice system can be attributed to significant corrections of higher-order harmonics in the lattice potential which are absent in the strict tight-binding limit. We calculate the dynamical consequences of the intermediate phase in detail to guide future experimental investigations for the observation of 1D SPME and the associated intermediate {(i.e., neither purely localized nor purely delocalized)} phase. We consider effects of interaction numerically, and conjecture the stability of SPME to weak interaction effects, thus leading to the exciting possibility of an experimentally viable nonergodic extended phase in interacting 1D optical lattices.
{Our work provides precise quantitative protocols for future optical lattice based experiments searching for mobility edges in one dimensional bichromatic incommensurate lattices, both in noninteracting and interacting systems.}
\end{abstract}

\maketitle

\section{Background}
{Ever since the classic work of Anderson 60 years ago~\cite{Anderson1958}, entitled ``Absence of diffusion in certain random lattices'', the study of quantum localization has been a central paradigm in condensed matter physics. Without any question, quantum localization is among the most-studied topics in physics encompassing systems as disparate as electrons in solids, nonelectronic (e.g., spins, phonons, plasmons, photons) excitations in solids, light, sound, and, most recently, cold atomic gases~\footnote{A search of the APS journal database with the word ``localization'' in title/abstract query returns more than 8000 articles.}. 
The subject has many facets and many aspects, but certain themes have attracted a great deal of attention recently, mainly because of the novel possibility of experimental investigations using precisely designed cold atomic systems. Indeed, Anderson localization in disordered systems has been studied extensively recently in cold atomic gases, both in lattice and continuous systems in the presence of controlled artificial disorder~\cite{Aspect2008Nature,Roati2008anderson,Aspect2010PRL,DeMarco2011Science,Aspect2012NaturePhysics,DeMarco2013PRL}. 
The nature of Anderson localization depends crucially on the dimensionality with two being the critical dimension~\cite{Abrahams1979,Lee1985RMP}, and much of the experimental activity has focused on one-dimensional systems where a well-known theorem asserts that all states are exponentially localized in the presence of any finite disorder (i.e., any noninteracting 1D quantum system is always localized in the presence of any disorder although the localization length could surpass the system size for weak disorder, making the system behave like a delocalized system for almost all practical purposes).  
A closely related problem, often referred to as the Aubry-Andre (AA) model (or more precisely, Aubry-Andre-Azbel-Harper model), has also been extensively studied in cold atomic gases, and the two localization problems (i.e., Anderson and AA) are often conflated together in the literature with both being called the Anderson localization phenomena with the basic issue in both problems being the question of whether the single-particle states are localized or delocalized in the system.

In the AA model, which is the nearest-neighbor tight-binding limit of a general two-potential incommensurate bichromatic lattice model [see Eq.~\eqref{Eq:IncommensuratePotential} below], a sharp localization transition occurs as the secondary potential strength is raised with all states being localized (delocalized) when the secondary potential  is larger (smaller) than a critical value determined by the nearest-neighbor hopping strength in the primary lattice.  
The strength of the secondary potential in the incommensurate AA model is often referred to as the ``disorder strength'' and the localization transition even in the AA model is often also called Anderson localization although the two localization phenomena are fundamentally different.
In particular, the localization transition in the AA model is not driven by quantum interference as in the Anderson model---instead, the AA localization in an incommensurate potential arises from the spectral properties of the \schrodinger{} itself.  This leads to the well-known difference between the two localization transitions:  In the 1D Anderson model (i.e., a 1D nearest-neighbor tight-binding model with on-site random disorder potential) any disorder makes all states localized whereas in the 1D AA model (i.e., a 1D nearest-neighbor tight-binding model with an incommensurate secondary potential) all states are localized or delocalized depending on the strength of the incommensurate potential.  
We note that neither model possesses a mobility edge in 1D, i.e., a critical energy separating localized and delocalized {energy eigenstates} (with all states being localized or delocalized in both cases depending on the ``disorder'' strength) although the Anderson model is known to have mobility edges in three dimensions (since two is the critical dimensionality for the model). 
In a practical sense, however, Anderson and AA models are quite similar---both are nearest-neighbor tight-binding models with an additional on-site potential term, which is either random (Anderson) or quasiperiodic with respect to the lattice (AA), and both have either all localized or all extended states (with no SPME) except that in the Anderson (AA) model the critical disorder strength necessary for causing localization is zero (finite).  Note that the on-site quasiperiodic potential (incommensurate with the periodicity arising from the primary lattice) in the AA model can be construed as an effective random disorder since it varies aperiodically from site to site albeit in a well-defined (rather than a random) manner. 

In the current work, we establish that any {realistic AA system} (i.e., any such bichromatic 1D system with two lattice potentials incommensurate with respect to each other), which is studied in the laboratory optical lattices, must necessarily manifest a single-particle-mobility-edge (SPME) separating delocalized and localized states although in the extreme tight-binding limit of the primary lattice potential being very deep, the system asymptotically approaches the AA limit with all states being delocalized {or localized} depending on whether the secondary lattice potential is below {or above} the critical value in the AA model.  
Perhaps more importantly from the experimental perspective, we show that an essential consequence of such an SPME is that 1D bichromatic incommensurate system manifests a generic intermediate (or a mixed) phase which is neither completely localized nor completely delocalized with observable signatures differing from either purely localized or purely extended phases.  
This is in striking contrast with the corresponding 1D Anderson model with disorder which does not have an SPME or an intermediate phase with all single-particle states being localized in the presence of even infinitesimal disorder.  
Thus, the bichromatic incommensurate lattice indeed is qualitatively different from disordered Anderson model   
{although in the tight-binding limit they behave similarly, because in that limit the bichromatic incommensurate lattice model reduces to the AA model. 
We note that earlier work has indicated the existence of SPMEs in various artificially created incommensurate 1D models~\cite{Economou1982,DasSarma1986PRL,DasSarma1988PRL,Fishman1988PRL,Thouless1988PRL,DasSarma1990PRB,Johansson1991,Biddle2009PRA,Biddle2010PRL,Biddle2011PRB,DasSarma2015PRL} but our work is the first one to establish the existence of SPMEs in the extensively experimentally studied (in cold atomic systems) bichromatic incommensurate potentials, which were always thought to be well-described by the tight-binding AA model. 

Perhaps an even more interesting question is the stability of the SPME and the intermediate phase in 1D incommensurate lattices to finite inter-particle interactions, i.e., the issue of currently active many-body-localization (MBL) phenomena as relevant to incommensurate bichromatic systems with SPMEs. 
There are two closely related, but distinct, questions here.  First, does such a system manifest MBL at all?  
Second, does the intermediate phase survive interaction {(or equivalently, is there a many-body mobility edge in the interacting system)}?  
We address both questions and answer them {in the affirmative} (at least, in small systems) by carrying out small system exact diagonalization.  There has been earlier work predicting the existence of an intermediate MBL phase in 1D incommensurate systems~\cite{Xiaopeng2015PRL,Mukerjee2015PRL,Xiaopeng2016PRB,Modak2016LIOM}, {leading to a possible manybody mobility edge}, albeit not in the simple bichromatic incommensurate potentials.  
Our work adds to this body of work, and our main new result is that SPME and intermediate phase exist in the extensively studied 1D bichromatic incommensurate potentials in spite of everybody assuming that the AA model without any SPME is the appropriate description for such incommensurate 1D lattices.  Thus, such SPME and the associated intermediate phase can now be experimentally studied in exactly the same systems where the physics of AA localization has already been studied~\cite{Roati2008anderson,Schreiber2015}. 
We propose very specific experimental protocols to search for the SPME and the intermediate phase in 1D cold atomic bichromatic incommensurate optical lattices. 
{Such experimental studies are crucial because whether our finding of an MBL through an exact diagonalization study of  small interacting systems is evidence in favor of the elusive many body mobility edge remains an open question, and much larger system numerics and/or experimental investigation would be necessary to decide this issue conclusively.}

\section{Introduction}

With the rapid progress in the last decade, ultracold atomic gases confined in optical lattices have now reached an era to simulate quantum many-body physics of lattice Hamiltonians~\cite{Zoller1998PRL,Lukin2002PRL,Lewenstein2007AdvPhys,Bloch2008RMP,Esslinger2010Review,Dutta2015Review,Liu2016Review}, 
exploring its inherent quantum supremacy, i.e., beyond the digital simulation capability of classical computers.  Both equilibrium quantum phase transitions and non-equilibrium many-body dynamics of ultracold atoms have been widely studied in the experiments. Quantum phases and phase transitions in strongly correlated Hubbard-type models have been experimentally demonstrated with bosonic~\cite{Greiner2002Nature,Folling2005Nature} and fermionic atoms~\cite{Jordens2008Nature,Schneider2008Science,Greif2013Science,Hart2015Nature}.
For example, the phase diagram of the Bose-Hubbard model has been experimentally mapped out. 
For the Fermi-Hubbard model, the long-sought doped antiferromagnetic phase has recently been experimentally confirmed~\cite{Mazurenko2016arXiv}. In recent years, quantum dynamics is attracting considerable interest, especially in understanding the fundamental questions of quantum thermalization and many-body localization, i.e., whether an interacting quantum system is generically ergodic or could sometimes become nonergodic by virtue of a localization transition in its Hilbert space~\cite{Schreiber2015,Kondov2015PRL,Bloch2016PRL,luschen2016evidence,Bloch2016Science,Smith2016NatPhys,YangLe2016PRA,Barnes2016PRB}.
The current work addresses the issue of both noninteracting and interacting MBL in 1D bichromatic incommensurate systems where fermions are moving in a system with two mutually-incommensurate periodic potentials so that the full system is aperiodic (but not random). 

Theoretically it has recently been shown that the  presence of interaction and strong disorder will lead to many-body localization in closed quantum systems where quantum ergodicity, i.e., the eigenstate thermalization hypothesis (ETH)~\cite{Deutsch1991PRA,Srednicki1994PRE} breaks down. The existence of MBL has now been  established through a perturbative calculation~\cite{BAA2006MBL}, extensive numerical simulations~\cite{Huse2007PRB,Peter2008PRB,Huse2010PRB,Moore2012PRL,Huse2013PRB,Pollmann2014PRL,Xiaopeng2015PRL,Mukerjee2015PRL,Luitz2015PRB,Laumann2015PRB,Moure2015EPL}, and a rigorous proof within some assumptions~\cite{Imbrie2016Proof}. The phenomena of MBL have been extensively explored in connection with quantum information~\cite{Bauer2013arealaw,Altman2015PRX,Huse2017ANDP} 
and local integrability~\cite{Abanin2013PRL,Huse2014PRBLIOM,Abanin2015PRBLIOM,Ros2015LIOM,Modak2016LIOM}, but the question of whether MBL is a generic phenomenon in isolated interacting quantum systems is still open. 
For example, the critical disorder strength defining MBL transition seems to increase with the simulation system size~\cite{Singh2015PRL}. 
The physics deep in the MBL phase has now been well studied. However understanding the transition from MBL to ergodic phases is extremely challenging and so far a reliable theoretical framework is still lacking.
Therefore, experiments in ultracold optical lattices incorporating both controlled disorder and designed interaction are playing a key role in elucidating the MBL physics since such analog quantum emulations can typically use systems much larger than the ones digital numerical simulations can address because of the exponential increase of the Hilbert space size with increasing system size.  

To shed light on the physics of MBL-to-ergodic transition, a controllable quantum system of ultracold atoms confined in a one-dimensional incommensurate optical lattice has been experimentally implemented~\cite{Schreiber2015}. 
Quantum dynamics of an out-of-equilibrium density wave state is found to exhibit nonergodic and thermalizing behaviors at the strong and weak incommensurate potential strengths, respectively.  
For a deep optical lattice, the lowest band is expected to be modeled with the tight-binding approximation, which leads to the well-known AA model~\cite{Roati2008anderson,Schreiber2015}. 
In the absence of electron-electron interactions, this model has a direct localization transition as we increase the incommensurate potential strength. 
The experimental system is then used to demonstrate the MBL-to-ergodic transition of the AA model by turning on a controlled interaction in the system~\cite{Schreiber2015}. 
Both noninteracting~\cite{Roati2008anderson} and interacting~\cite{Schreiber2015} localization has been experimentally studied in the AA model in the tight-binding limit of the deep primary lattice potential.  
The current work addresses the situation where the primary lattice potential is not necessarily deep so that AA tight-binding limit approximation may break down. 

In this work we study the incommensurate optical lattice realized in the experiment~\cite{Schreiber2015}, taking into account continuous degrees of freedom beyond the tight-binding approximation---we note that in the general bichromatic incommensurate potential situation, a tight-binding lattice approximation does not apply since the system is aperiodic and the full two-potential \schrodinger{} must be solved exactly (which we refer to as the ``continuous'' case since both primary and secondary potentials are now treated equivalently spatially with the quantum hopping on the primary lattice no longer playing any special role).
Our theory uncovers the existence of an SPME in the continuous incommensurate lattice, not captured by the AA model description. 
We find that, unlike the AA model, which has a direct localization transition point, the continuous system has an intermediate region between fully localized and delocalized phases which supports SPME. Our results imply the necessity of incorporating mobility edge physics for many-body localization of cold atoms in incommensurate optical lattices since the applicability of a tight-binding approximation cannot be determined \emph{a priori} (and the localization physics in the continuous and the tight-binding situations are qualitatively different). 
This makes the issue of how localized and extended degrees of freedom interact~\cite{Rahul2014PRB,Xiaopeng2015PRL,Mukerjee2015PRL,Rahul2015PRB,Xiaopeng2016PRB,Roeck2016PRB,Bauer2017PRB,gornyi2017absence} unavoidable in understanding the MBL transition for  such systems. 
Our study indicates that the behavior of MBL (or even single-particle-localization) physics could be drastically different in continuum and lattice Hamiltonians (nominally representing the same problem), as has recently been discussed in the context of Anderson localization in disordered systems~\cite{Rahul2014PRB,Roeck2016PRB,gornyi2017absence}.  
We note that our work now allows this important issue (i.e., the interplay between MBL and SPME) to be studied in ultracold optical lattices by starting with a noninteracting bichromatic incommensurate potential system which manifestly has an SPME.


To guide future optical lattice experiments in the 1D bichromatic incommensurate potential system, we  evaluate the width of the intermediate phase in the \iLM{}. We find that for practical and easily accessible experimental values of the incommensurate lattice potential, the intermediate phase always has a considerable width, which continues to increase as the primary lattice depth is lowered (In the limit of a very deep primary lattice potential, we asymptotically reach the AA limit of vanishing intermediate phase with the system being either completely localized or completely extended depending on the strength of the secondary potential.). 
We propose two physical observables, i.e., density imbalance $\mI$ and edge density fraction $\mD$, that can be combined to precisely define this intermediate phase in the experiment. 
In particular, we establish that the localized (extended) phase has finite values of $\mI$ $(\mD)$ whereas the intermediate phase has both $\mI$ and $\mD$ finite, distinguishing it from either the purely localized (finite $\mI$, but vanishing $\mD$) or the purely delocalized (finite $\mD$, but vanishing $\mI$) phase. 

Having established the existence of an intermediate phase in the \iLM{} due to SPME, we introduce a conceptually transparent two-band toy model to show that the essence of this intermediate phase is the coexistence (or a mixture) of delocalized and localized eigenstates at different energies. Specifically, we use two copies of the AA model with different hopping energies to construct an effective two-band model that manifestly has an SPME by definition. We show that every aspect of the intermediate phase in the \iLM{} can be simply understood in this two-band model. 
In principle, this two-band model manifesting SPME can also be studied in experimental optical lattices, but it may be much easier to use the bichromatic incommensurate model for the experimental search of SPME since such 1D lattices have already been studied in the laboratory.

Although it is difficult to theoretically study the interaction effects on this intermediate phase in the thermodynamic limit, we carry out small system exact diagonalization studies for both models (bichromatic incommensurate potential and simple two-band AA), and our numerical results indicate the survival of the intermediate phase in small interacting systems. We thus provide a strong hint that the intermediate phase and SPME are stable to finite interactions, and MBL exists even in the presence of an SPME, but whether this conclusion is valid in the thermodynamic limit remains an important question necessitating further investigations. 
Actual experiments in optical lattices involving hundreds of atoms could further strengthen the existence (or not) of MBL in interacting bichromatic incommensurate systems. Our work, however, definitively establishes the existence of the SPME and the associated intermediate phase in the noninteracting incommensurate potential model, and this could be easily verified in already existing systems by measuring the imbalance parameters $\mI$ and $\mD$ as mentioned above.

Our paper is organized as follows. In Section~\ref{Section:iLM} we introduce the bichromatic \iLM{}, and discuss in detail its SPME and the associated intermediate phase. In Section~\ref{Section:ExpSignature} we propose and calculate two physical observables that can be used to diagnose the existence of the intermediate phase. We show that the generic \iLM{} behaves qualitatively differently from the Aubry-Andre tight-binding limit, and thus it is possible to distinguish them experimentally. In Section~\ref{Section:spModel} we construct a two-band toy model to demonstrate that the essence of the \iLM{} is the coexistence of localized and delocalized eigenstates. In Section~\ref{Section:Summary} we present some additional discussions and summarize our findings. 
Some of the detailed technical aspects of our theoretical results are presented in the Appendix in order to keep the main presentation smooth without unnecessary technical details and easy to follow. 

\section{The \iLM{} \label{Section:iLM}}

The single-particle Hamiltonian of the 1D bichromatic incommensurate potential in the continuum limit reads 
\begin{align}
	\mH(x) = -\dfrac{\hbar^2}{2M}\dfrac{d^2}{dx^2}+\dfrac{V_s}{2}\cos(2kx)+\dfrac{V_d}{2}\cos(2k\alpha x +\phi), \label{Eq:IncommensuratePotential}
\end{align}
where $\alpha$ by definition (in the theory) is an irrational number indicating that the system overall is aperiodic since the one-particle potential [i.e., the sum of the last two terms in Eq.~\eqref{Eq:IncommensuratePotential}] never repeats itself as the primary (indicated by the $V_s$ term) and the secondary (indicated by the $V_d$ term) potentials are incommensurate with respect to each other. 
Throughout our discussions we will use the lattice constant $a = \pi/k$ as the length unit and the photon recoil energy of the \emph{primary} lattice $E_{R} = \hbar^2k^2/2M$ as the energy unit. (These definitions are closely tied to our {intention to connect} Eq.~\eqref{Eq:IncommensuratePotential} to the ultracold optical lattice systems used experimentally to study the \iLM{}.)  
Note that because of the periodic boundary condition, we cannot use an actual irrational ratio $\alpha$. 
(This is of course also true in the experimental situation where $\alpha$ is always a rational number chosen appropriately to mimic an irrational for the given system size.) 
Instead, we choose to approximate $\alpha$ by the ratio of two adjacent Fibonacci numbers $F_{N-1}/F_N$, with $F_{N-1}$ and $F_N$ being the period of the primary and secondary lattice, respectively, which approaches the Golden Ratio $(\sqrt{5}-1)/2$ in the thermodynamic limit. 
Note that the only thing we must ensure is that $\alpha$ is such that the system remains aperiodic within the system size of our simulations. 
We solve the above Hamiltonian in a system with length $L=610$ by discretizing the real-space coordinates and diagonalizing the resulting matrix. 
{Our system size is chosen with experimental optical lattices in mind, and our results and conclusions do not change at all if we increase the system size with the only difference being that the localization transition becomes sharper with increasing size.} 
We make sure that the discretization used for our exact diagonalization is on a sufficiently fine mesh so as to ensure convergence of all results.
{We also mention that we do not consider the effect of a background trap potential in Eq.~\eqref{Eq:IncommensuratePotential} (or in the rest of the main text) because it is possible to have very flat background confining potentials in optical lattices where our proposed experiments should be performed.  The existence of a smooth trap potential will simply broaden all the transitions in the system, making it difficult to discern localized, intermediate, and extended phases.  We do, however, discuss the effect of a smooth trap potential in Appendix~\ref{Section:Trap} for the sake of completeness.
} 

\begin{figure}[!]
\includegraphics[scale=1]{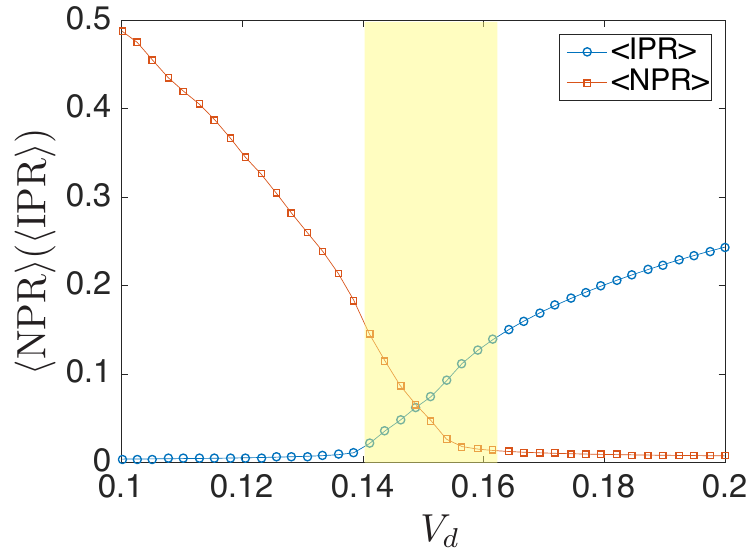}
\caption{\label{Fig:NPRandIPR} Averaged NPR and IPR for all eigenstates in the lowest band in the \iLM{} [Eq.~\eqref{Eq:IncommensuratePotential}]. Here the depth of the primary lattice potential is $V_s = 8$, $\alpha = 377/610$, and the system size is $L=610$. The shaded region is where non-ergodic metallic behavior is likely to appear in this model.
Similar results are obtained for other values of $V_s$ with the shaded regime increasing (decreasing) in width with decreasing (increasing) $V_s$.
}
\end{figure}

{One unique property of this model defined by Eq.~\eqref{Eq:IncommensuratePotential} is that it exhibits an SPME~\cite{PhysRevA.64.033416,PhysRevA.75.063404}.
One way to show this is to calculate the inverse participation ratio (IPR) of all eigenstates in the lowest band. Such a quantity is designed to characterize the localization properties of an eigenstate.} 
In particular, the IPR for the $i$th eigenstate $u_m^{(i)}$ is defined as 
\begin{align}
	\text{IPR}^{(i)} = \dfrac{\sum_{m}\abs{u_m^{(i)}}^4}{\left(\sum_{m}\abs{u_m^{(i)}}^2\right)^2}, 
\end{align}
where $m$ labels the real space (discretized) coordinates. 
It vanishes for spatially extended states while {remaining} finite for localized states. 
There also exists a complementary quantity, the normalized participation ratio (NPR), 
\begin{align}
	\text{NPR}^{(i)} = \left[L\sum_{m}\abs{u_{m}^{(i)}}^4\right]^{-1}, 
\end{align}
which, on the contrary, remains finite for spatially extended states but vanishes for localized states. 
Direct numerical calculations of IPR (finite/zero for localized/extended states) and NPR (zero/finite for localized/extended states) therefore enable identifications of individual energy eigenstates of the system as being localized or delocalized~{\footnote{Note that this statement is exact only in the thermodynamic limit. For a finite 1D system, IPR goes like $L^{-2}$ for extended states, while NPR goes like $L^{-1}$ for localized states, where $L$ is the size of the system. }}.
{We note that for a single eigenstate IRP and NPR are trivially connected, but an averaging over all states (as we do) leads to averaged IPR and NPR providing complementary information for extended and localized states.} 

We calculate the IPR and NPR for all eigenstates in the lowest band of this model, and their average values are shown in Fig.~\ref{Fig:NPRandIPR}. (For an energy-resolved plot of IPR we refer to Appendix~\ref{Appendix:SPME}.) 
We can see that depending on the strength of the incommensurate potential $V_d$, there exist three distinct phases for a given strength of the primary lattice potential $V_s$. 
For a small $V_d~(<0.14)$, all eigenstates in the lowest band remain extended, as indicated by a vanishing $\average{\text{IPR}}$ (where $\average{\cdot}$ denotes an average value). This regime is adiabatically connected to the $V_d=0$ limit, where all eigenstates are delocalized.  
For a sufficiently large $V_d~(>0.16)$, all eigenstates are localized, as suggested by a vanishing $\average{\text{NPR}}$, which arises when the depth of the incommensurate lattice potential overcomes the bandwidth of the lowest band. Within the AA model description of Eq.~\eqref{Eq:IncommensuratePotential}, where {$V_s$ (i.e., how deep it is) defines the nearest-neighbor tight-binding model}, 
all one expects are {the finite (zero) regimes of NPR (IPR) defining extended (localized) spectra}
{with the {sharp} AA transition defined by a finite strength of $V_d$}---we do not anticipate an intermediate regime with both IPR and NPR being finite within the AA physics which does not manifest any SPME.
However, an intermediate regime ($0.14<V_d<0.16$ in Fig.~\ref{Fig:NPRandIPR} for $V_s = 8$) clearly exists in this \iLM{}, in which both $\average{\text{IPR}}$ and $\average{\text{NPR}}$ remain finite, indicating that the spectrum of Eq.~\eqref{Eq:IncommensuratePotential} allows for a phase which has both spatially extended and localized states as its eigenstates (at different energies){,} in contrast to the predictions of the AA model. 
The width of this regime is controlled by the bandwidth of the lowest band, and will be reduced exponentially as we approach the deep lattice limit (when $V_s$ is large). 
This intermediate regime is characterized by the coexistence of localized and extended states at different energies (with the SPME separating them {see the next section}), and is the focus of this study. In particular, as we show below, this intermediate regime is  characterized by a non-ergodic metallic behavior, where the system is delocalized but nonthermal. 
This is perhaps the simplest (and the most experimentally viable) example of a nonergodic metallic phase~\cite{Xiaopeng2015PRL,Xiaopeng2016PRB} in the context of widely studied cold atom optical lattices since Eq.~\eqref{Eq:IncommensuratePotential} has already been implemented in cold atomic systems. 

{
\begin{widetext}
\section{SPME and fractal bandstructure in the intermediate phase \label{Appendix:SPME}}
\begin{figure*}[h]
\includegraphics[scale=0.85]{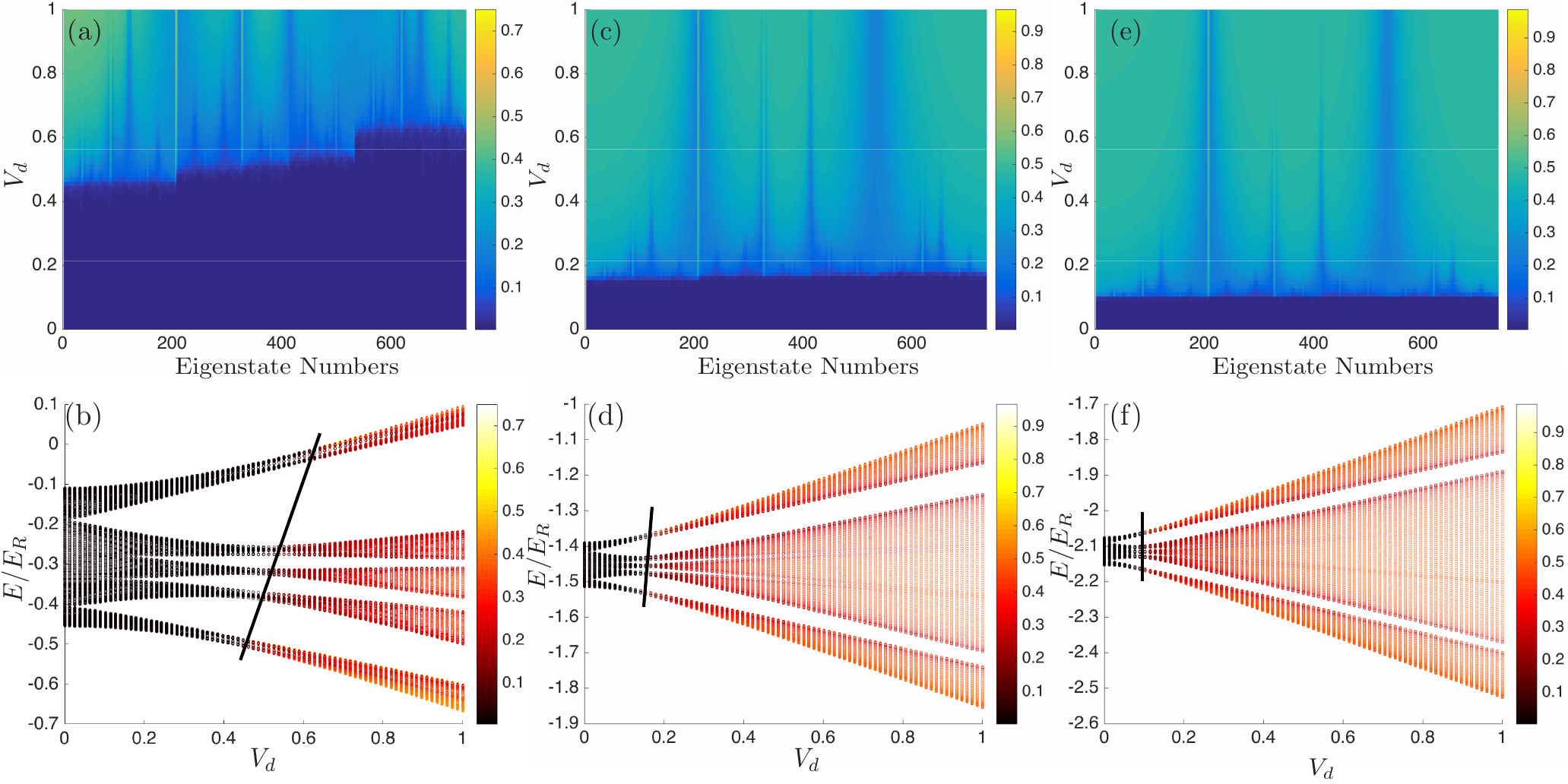}
\caption{Inverse participation ratio (IPR) of all eigenstates in the lowest band of the \iLM{} for $\alpha = 532/738$ and a primary lattice depth of $V_s=4$ [(a)-(b)], $V_s = 8$ [(c)-(d)], and $V_s = 10$ [(e)-(f)] respectively. The plots in the first row show IPR simply as a function of eigenstate numbers, while those on the second row are further augmented by the energy of each eigenstate. The solid lines in the second row serve as a guide to the eye, and mark out the single-particle mobility edge in each figure. \label{Fig:SPME}}
\end{figure*}

It is also instructive to understand the intermediate phase from the structure of the energy spectrum, particularly since the intermediate phase is in some sense a critical phase in between the fully localized and the fully extended phase
We have mentioned that the intermediate phase in the \iLM{} arises from a coexistence between localized and delocalized states. 
Such a structure is revealed explicitly in Fig.~\ref{Fig:SPME},  which shows the IPR of individual eigenstates in the lowest band of this model. 
A sharp distinction is seen between deep (e.g., $V_s = 10$) and shallow (e.g., $V_s = 4$) lattices: for deep lattices, the localization properties are well approximated by the \AAM{} model, with all localization transitions happening at the same energy; in contrast, the localization transition is clearly energy dependent in shallow lattices. The existence of an SPME is the ultimate reason for the intermediate phase in the \iLM{}.  

One prominent feature in Fig.~\ref{Fig:SPME} is that the continuum energy spectrum at $V_d = 0$ gets split into multiple subbands as $V_d$ is turned on. Such a structure is revealed more clearly in the density-of-states plot in Fig.~\ref{Fig:DOS}, where a series of gaps emerges in the spectrum when $V_d > 0$. 
However, note that our discussions are entirely focused on a single band---the lowest band of the original periodic system. The gaps that appear in the spectrum only separate the minibands arising due to the disorder potential (i.e., the secondary potential in the bichromatic lattice). 
Therefore, the mobility edge that appears in the spectrum still lies in the middle of the band arising from the primary lattice---the mini-gaps are connected with the `disorder' potential, not the primary potential.
\end{widetext}

\begin{figure}[!]
\includegraphics[scale=0.9]{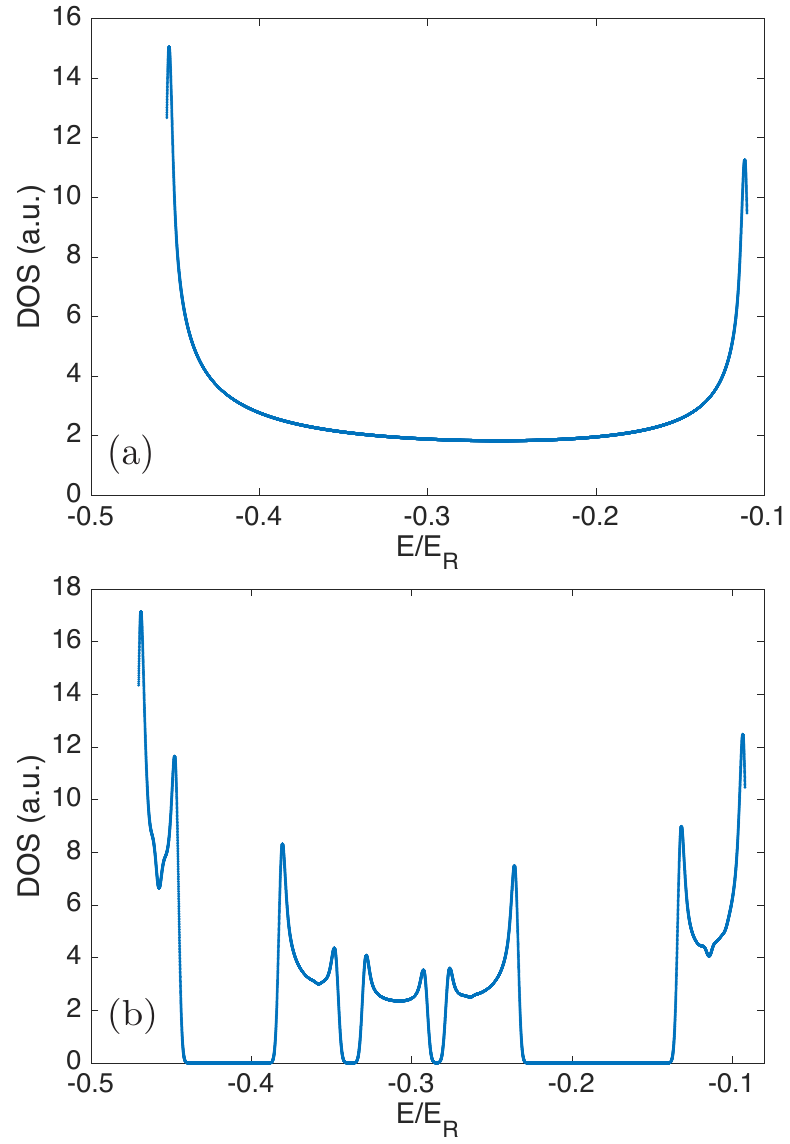}
\caption{\label{Fig:DOS} Density of states for the \iLM{}, with $V_s = 4$ and $\alpha = 532/738$. (a) and (b) corresponds to $V_d = 0$ and $V_d = 0.25$, respectively. The subband structure is clearly seen in (b). }
\end{figure}

\begin{figure}[!]
\includegraphics[scale=1]{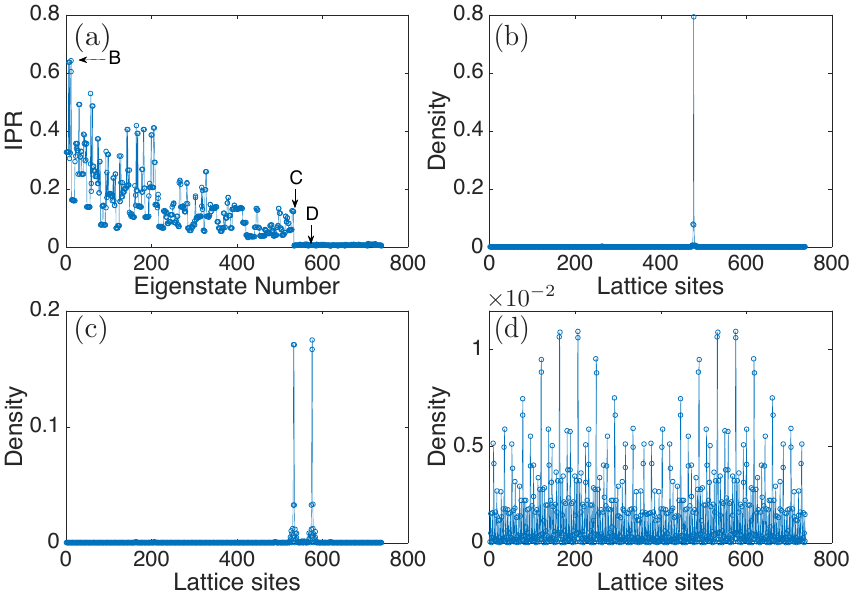}
\caption{\label{Fig:IPR} (a) IPR of all eigenstates in the lowest band of the \iLM{} with $\alpha = 532/738$, $V_s = 4$ and $V_d = 0.55$. 
The particle density of the three states labeled by `B', `C', and `D' in (a) are correspondingly shown in (b)-(d). 
}
\end{figure}

Having examined the general structure of the energy spectrum of the \iLM{}, we now turn to its intermediate phase, which has many interesting properties. 
We first study the spatial spread of the wavefunction for states in the intermediate phase. 
Fig.~\ref{Fig:IPR} shows the IPR of all eigenstates in the lowest band when $V_d = 0.55$ (which is in the intermediate phase), as well as three typical particle density distributions. 
In particular, state `C' in Fig.~\ref{Fig:IPR}(a) is the critical state that separates localized and extended states. 
One hallmark of such a critical state is its spatial density distribution [see Fig.~\ref{Fig:IPR}(b)]: it looks like a localized state in a subregion, but there is a nonzero probability to find additional density peaks throughout the system. Such a property is distinct from both localized states [Fig.~\ref{Fig:IPR}(b)] and extended states [Fig.~\ref{Fig:IPR}(d)]. 
This is a direct manifestation of the singular continuous spectrum associated with the critical state at the SPME whereas the localized and the extended states have point and continuous spectra, respectively. 

\begin{figure}[!]
\includegraphics[scale=1]{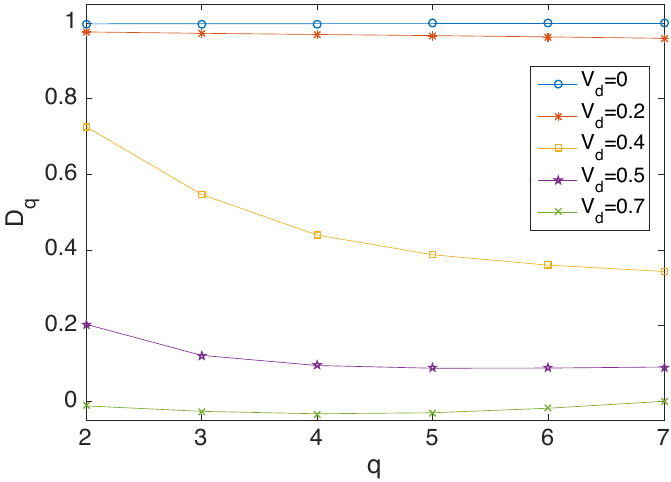}
\caption{\label{Fig:MultiFractal} Fractal dimensions of the wave functions in the \iLM{}. In this figure $V_s = 4$ and $\alpha = 532/738$. }
\end{figure}

The peculiar spatial density distribution for states in the  intermediate phase is closely related to their multifractality, which is a common feature of states at the mobility edges for Anderson transitions~\cite{RMP_Anderson}. 
The multifractal structure can be captured by the (generalized) IPR~\cite{RMP_Anderson}, defined as 
\begin{align}
	P_q = \int dx \,\abs{\psi(x)}^{2q}. 
\end{align}
At criticality (i.e., at the mobility edge), $P_q$ show an anomalous scaling with the system size $L$, $\average{P_q}\sim L^{-\tau_q}$. This continuous set of exponents $\tau_q$ captures the critical behavior of the wave function. 
It is common to introduce fractal dimensions $D_q$ via $\tau_q = D_q(q-1)$. For one-dimensional systems $D_q = 1$ in a metal (continuous spectrum for extended states) and $D_q = 0$ in an insulator (point spectrum for localized states), while in the intermediate phase $D_q$ is a nontrivial function of $q$, which is a manifestation of the multifractal structure of the wave function. 
Figure~\ref{Fig:MultiFractal} shows a plot of our calculated $D_q$ for various different values of $V_d$ in the bicromatic incommensurate model. 
The $V_d=0$ curve has a constant $D_q = 1$, indicating a metallic system. 
The $V_d = 0.2$ curve  shows a small but finite departure from $D_q = 1$ despite the fact that the system remains extended. This should be due to finite size effects in the scaling analysis. 
In the opposite limit of large $V_d$ ($V_d=0.7$), we have $D_q\simeq 0$, typical for a completely localized one-dimensional system. The weak dependence on $q$ is again a result of finite-size effects. 
For intermediate values of $V_d$, however, $D_q$ develops a nontrivial $q$ dependence, reflecting the multifractal structure in the eigenstate wave functions. 
Thus, the intermediate phase is critical manifesting multifractality. 

\begin{figure}
\includegraphics[scale=1]{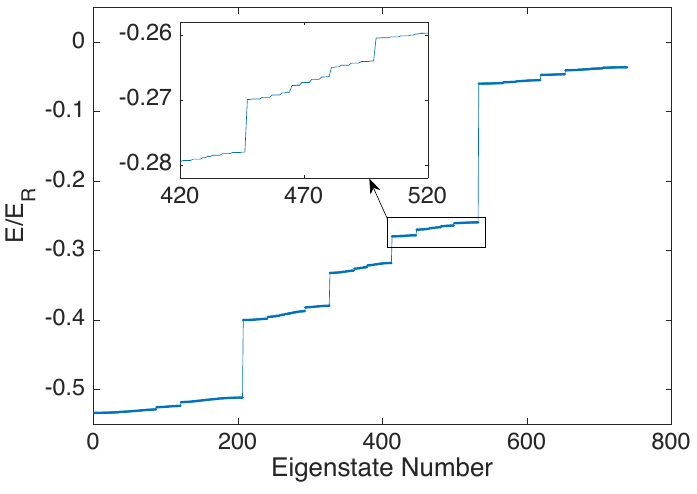}
\caption{\label{Fig:Fractal} Fractal and self-similarity structure of the energy spectrum in the intermediate phase. }
\end{figure}

Another interesting aspect of the intermediate phase is the fractal structure of the energy spectrum, as shown in Fig.~\ref{Fig:Fractal}. In particular, the fine structure of a subband carries the same (fractal) structure of the entire energy band. Such a self-similarity structure is ubiquitous in systems with a fractal energy spectrum~\cite{Hofstadter,Kohmoto1983,Kohmoto1986,Economou1986,RandomWalk2010}. 
}

\section{Experimental signatures of the single-particle mobility edge \label{Section:ExpSignature}}
Having established the existence of three phases (i.e., extended for smaller $V_d$, localized for larger $V_d$, and intermediate for intermediate $V_d$) {in} the model defined by Eq.~\eqref{Eq:IncommensuratePotential}, we now discuss how they can be identified in the experiment using optical lattices. We will propose two complementary experimental observables that can be used to {directly distinguish} the three phases. 
{These two physical observables correspond roughly to IPR and NPR discussed above}, which, by themselves, are not directly experimentally accessible. 
Moreover, we show that by slightly modifying the existing experimental setup, the intermediate phase {becomes} much wider, which will facilitate its experimental observation. 
One important distinction to keep in mind in dealing with experimental situations is that often the experimentally measurable quantities are not direct eigenstate behaviors, and therefore, IPR, NPR, etc., which are various measures of direct energy eigenstate properties are not experimentally obtainable. 

\subsection[density]{Density imbalance \texorpdfstring{$\mI$}{I} and edge density fraction \texorpdfstring{$\mD$}{D}}

We first propose two experimental observables that can combine to serve as precise operational definitions for the three different phases. 
Both can be directly measured (and have been measured) in cold atom optical lattice experiments~\cite{Schreiber2015,Bloch2016Science}. 
The first observable is called the \emph{density imbalance} $\mI$, defined as the difference between particle densities on even and odd lattice sites: 
\begin{align}
	\mathcal{I} = \dfrac{n_\text{even}-n_\text{odd}}{n_\text{even}+n_\text{odd}}, 
\end{align}
which ranges from $+1$ (for all particles on even sites) to $-1$ (for all particles on odd sites). Without loss of generality we define the lattice sites as belonging to those defined by the primary potential $V_s$ although any other way of defining this lattice would not change anything in our consideration. 
Here $n_\text{even}$ ($n_\text{odd}$) denotes the total particle density on even (odd) lattice sites.  
To measure this quantity one should start from a charge density wave state with all particles loaded {into} odd sites only. 
The system is then evolved under the incommensurate lattice Hamiltonian in Eq.~\eqref{Eq:IncommensuratePotential} for a sufficiently long time until a stationary state is reached. 
We note that by definition the beginning charge density wave state is not an energy eigenstate of the Hamiltonian, but is a simple physical state  obtainable experimentally in optical lattices~\cite{Schreiber2015}. 
{As a result, the imbalance measurement intrinsically contains an average over all eigenstates in the lowest band.} 
{We expect (and verify, as described below) that in the long-time limit $\mI$ will vanish in a delocalized system, but remains finite in the presence of localized states.} 

The second observable we propose is the \emph{edge density fraction} $\mD$, defined as 
\begin{align}
	\mathcal{D} = \dfrac{2n_\text{right}}{n_\text{right}+n_\text{left}}, 
\end{align}
which ranges from $0$ (for all particles in the left half of the system) to $2$ (for all particles in the right half of the system). 
Here $n_\text{left}$ ($n_\text{right}$) denotes the total particle density in the left (right) half of the system. 
The corresponding measurement consists of initializing all particles in the left half of the system and then measuring the fraction of the particles ending up in the right half of the system in the long-time limit. In particular, if all eigenstates in the system are delocalized (localized), we expect $\mD = 1$ ($\mD = 0$). 
Again, the situation to measure $\mD$ (i.e., starting with all particles in the left half of the system and then measuring the density fraction after a long time) {can be} achieved experimentally. 

\begin{figure*}[!]
\includegraphics[scale=1]{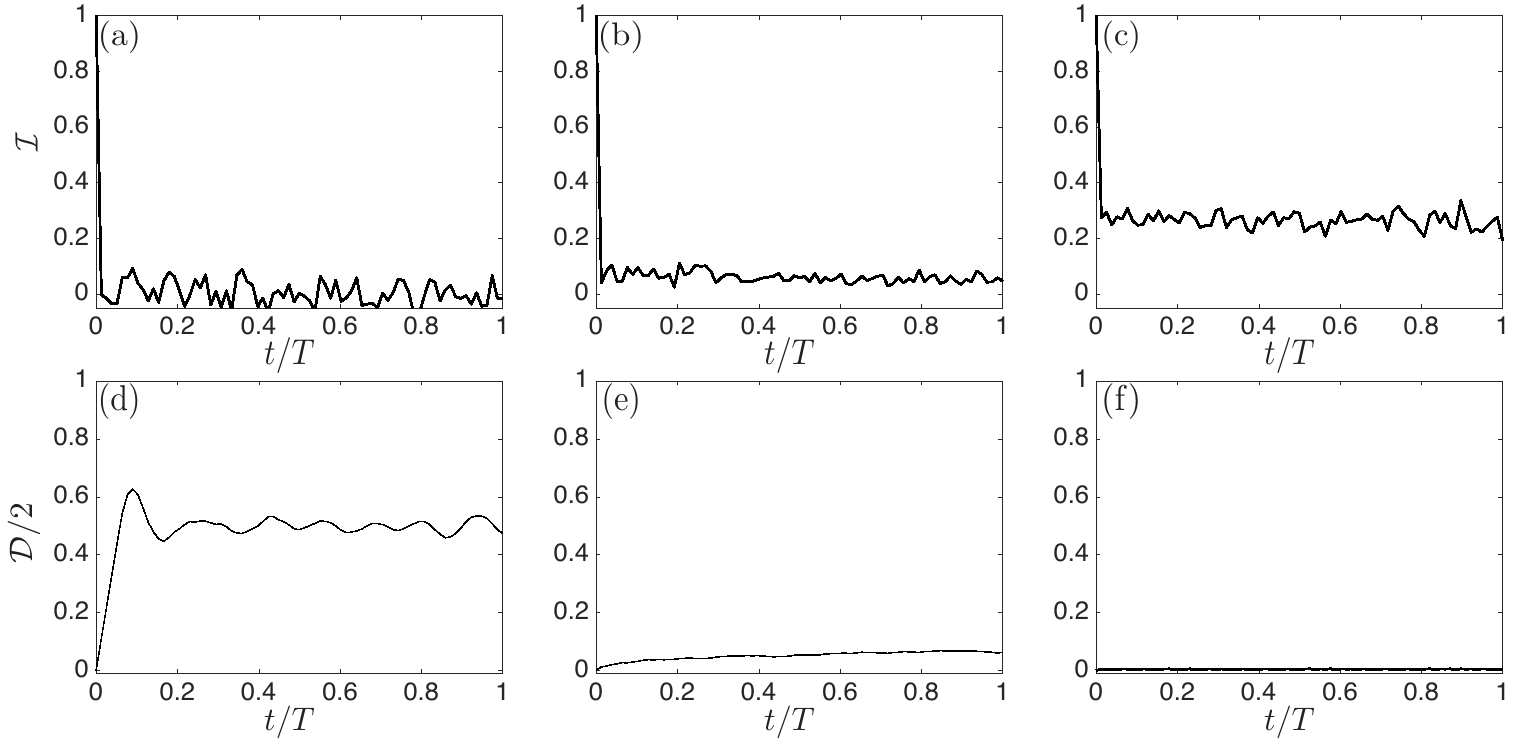}
\caption{\label{Fig:DensityImbalanceEvolution} 
Time evolution of the density imbalance $\mI$ and edge density fraction $\mD$ when the lowest band of the \iLM{} in Eq.~\eqref{Eq:IncommensuratePotential} is half filled.  
(a)-(c) show the density imbalance $\mI$ for $V_d=0.05$, $0.15$, and $0.20$, respectively, while (d)-(e) show $\mD$ for the same values of $V_d$. 
The total time is $T=10^5$ in units of $T_0 = 2m/(\hbar k^2)$, $\alpha = 532/738$, and the system size is $L=738$. The entire system is in the deep lattice limit ($V_s = 8$). 
According to Table~\ref{Table:thermalization}, (a) and (d) correspond to a delocalized and thermal phase, (b) and (e) correspond to a delocalized but nonthermal phase, while (c) and (e) correspond to a localized and nonthermal phase. 
}
\end{figure*} 

These two quantities capture different (and complementary) aspects of localization properties of a system, and thus can be combined to yield a complete understanding of the experimental localization transition (similar to what IPR and NPR achieve for the theory). 
Specifically, the index $\mI$ can diagnose whether \emph{all} eigenstates in the system are \emph{delocalized}, because it will become nonzero {as soon as some localized states} exist in the spectrum. 
In contrast, the index $\mD$ provides a complementary diagnosis by checking whether \emph{all} eigenstates in the system are \emph{localized}, because it will become nonzero as soon as delocalized states exist. 
As a result, when $\mI$ vanishes and $\mD$ remains finite, the system is in a delocalized phase. In contrast, when $\mD$ vanishes and $\mI$ remains finite, the system is in a localized phase. 
If both $\mD$ and $\mI$ {are} finite, the system is in a mixture of localized and delocalized states, which we identify as the intermediate phase between localized and delocalized phases arising from the existence of {an} SPME in the spectrum leading to a mixed state which is neither purely localized nor purely extended. 
{This is then the experimental signature for the existence of SPME and intermediate phase.} 
Table~\ref{Table:thermalization} summarizes the above discussions and clearly demonstrates how the three distinct phases are operationally defined. 
Very loosely speaking, $\mI$ and $\mD$ are qualitative measures of IPR and NPR respectively, and in the intermediate phase both should be finite.

\begin{table}[!]
\centering
\caption{\label{Table:thermalization} Localization vs thermalization properties in an \iLM{}.}
\begin{tabular}{l|cc}
\hline\hline
                           & $\mI$ & $\mD$ \\ \hline
Localized                  & finite        & 0       \\ \hline
Delocalized \& nonthermal & finite       & finite  \\ \hline
Delocalized \& thermal     & 0             & 1    \\ \hline\hline
\end{tabular}
\end{table}


\begin{figure*}[!]
\includegraphics[scale=1]{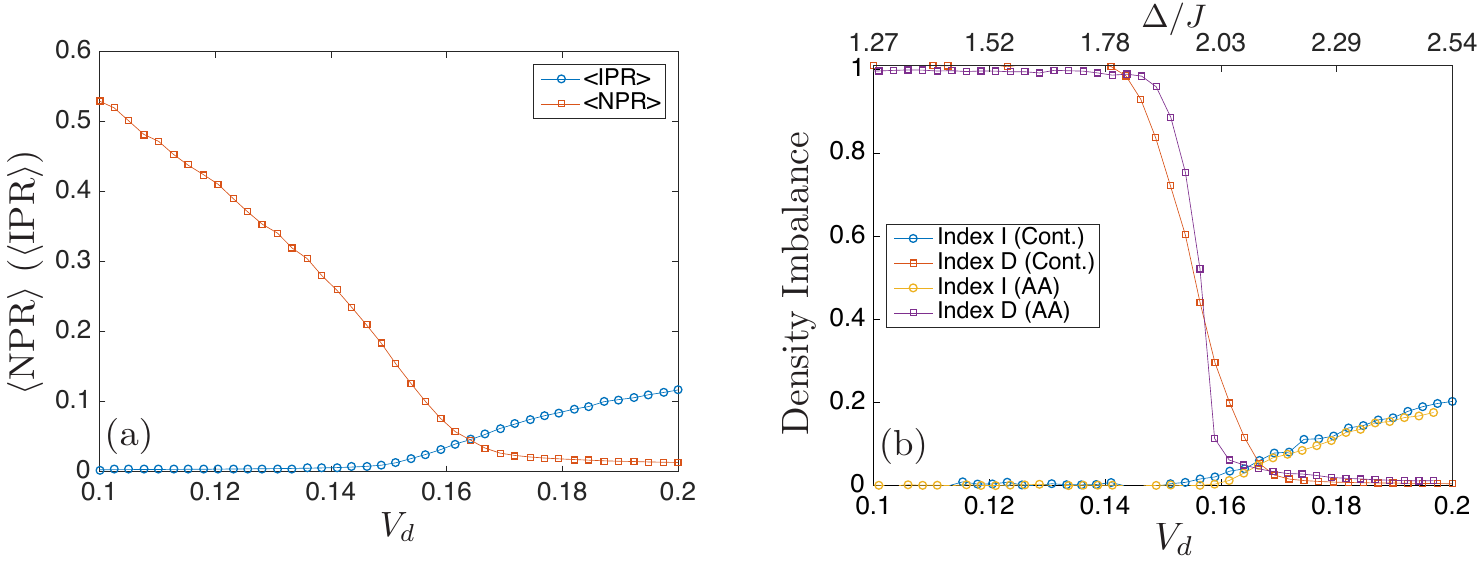}
\caption{\label{Fig:Imbalance-Vs8} 
(a) Averaged NPR and IPR and (b) density imbalance and edge density fraction for the \iLM{} with $V_s = 8$ and $\alpha = 532/738$ in a system with $L=738$ sites. Panel (b) also draws a comparison between the results from the continuum model and the \AAM{} model. 
The $x$-axis label in (b) on the top and bottom corresponds to the \iLM{} and the \AAM{} model results, respectively. 
}
\end{figure*} 

{ Here we propose a specific experimental dynamical protocol to probe the three phases found in the \iLM{}.   
We begin with $V_s = 8$ and $V_d = 0$ in Eq.~\eqref{Eq:IncommensuratePotential}, and make the lowest band half-filled. We then {adiabatically introduce} a new potential to create the desired initial states for $\mI$ and $\mD$. For example, for $\mI$ the initial charge density wave state is prepared by {adiabatically} introducing a new potential $V_2\cos[k(x-a/2)]$, which confines all particles onto even sites only. After the initialization, we {quickly} turn off this potential and {quickly} turn on the incommensurate lattice potential $V_d$. 
We then calculate the quench dynamics of the initial charge density wave state and evaluate $\mI$ at different times. The edge density fraction $\mD$ can be obtained in a similar way. 

We also note that a rational $\alpha = 532/738$ will be used in most of our calculations, which is the approximate $\alpha$ realized in recent optical lattice experiments~\cite{Schreiber2015}. 
Such a choice allows us to better guide future experiments. 
There is a caveat in doing this, as one should keep in mind that a rational $\alpha$ will always lead to a delocalized phase in the thermodynamic limit. Yet, in a finite system the localized phase in this model will persist, and the only effect of a rational $\alpha$ is to smooth out the localization transition~\cite{Wilkinson1984critical,Zhang2015Mathieu}.  
These features are well-understood already in the context of studying AA localization properties in optical lattices, and our analysis does not introduce any extra complications. 
} 

Figure~\ref{Fig:DensityImbalanceEvolution} shows a typical time evolution of $\mI$ and $\mD$ in a lattice with depth $V_s = 8$, which clearly demonstrates the three possible phases we propose. 
We can see that when $V_d = 0.05$, $\mI$ approaches zero [(a)] while $\mD$ approaches $1$ [(d)], which is a typical behavior for a completely delocalized system. 
In the opposite limit of $V_d = 0.20$, $\mI$ remains finite [(c)], while $\mD$ becomes vanishingly small [(f)], indicating a localized system. 
In the intermediate regime (e.g., $V_d = 0.15$), {however,} both $\mI$ and $\mD$ are finite, suggesting a nonergodic metallic behavior associated with the intermediate phase. 
Thus, these two physically measurable quantities can indeed differentiate the three phases in the experiment. 
{This figure also demonstrates clearly that the time scales needed by the imbalance and edge density fraction to reach their stationary values are significantly different, which is also a distinct experimental prediction of our theory.}
Figure~\ref{Fig:Imbalance-Vs8} shows a complete phase diagram by plotting {the stationary values} of $\mI$ and $\mD$ (as well as $\average{\text{NPR}}$ and $\average{\text{IPR}}$) as a function of the incommensurate potential strength $V_d$ for a fixed value of $V_s$ (other values of $V_s$ produce qualitatively similar phase diagrams with the intermediate phase being strongly suppressed with increasing $V_s$ as the system approaches the tight-binding AA limit). 
{We emphasize that Fig.~\ref{Fig:Imbalance-Vs8} explicitly demonstrates that $\mathcal{I}$ and $\mathcal{D}$ are perfectly acceptable physical variables defining the localized/extended phase diagram (including the intermediate phase) since the phase diagrams obtained from $\mathcal{I}$/$\mathcal{D}$ agree quantitatively with that obtained from $\average{\text{IPR}}$/$\average{\text{NPR}}$.} 
One can clearly observe the existence of an intermediate phase during the localization transition as $V_d$ is increased (and the system is driven towards a localized phase).

\subsection{Connection to the \AAM{} model}
It is widely believed that in the deep lattice limit of Eq.~\eqref{Eq:IncommensuratePotential} ($V_s\gg1$, with $V_s\gg V_d$), the physical properties of the \iLM{} can be studied with the well-known single-band tight-binding \AAM{} (AA) model, 
\begin{align}
	Eu_n = -J(u_{n-1}+u_{n+1}) + \Delta\cos(2\pi\alpha n+\phi) u_n, \label{Eq:AAModel} 
\end{align}
which contains only the nearest-neighbor hopping and on-site potential terms. 
Therefore, it has often been used to guide cold atom experiments studying localization physics emulating the Hamiltonian defined by Eq.~\eqref{Eq:IncommensuratePotential}~\cite{Roati2008anderson,Schreiber2015}. 
Such a tight-binding model has a self-duality point at $\Delta/J$\,$=$\,$2$, which separates the localized phase where all states are localized ($\Delta/J>2$) from the delocalized phase where all states are extended ($\Delta/J<2$). 
Moreover, it is possible to establish a mapping between the tight-binding and continuum model parameters in this limit. 
Specifically, the nearest-neighbor hopping $J$ is solely determined by the primary lattice depth $V_s$~\cite{Bloch2008RMP}, 
\begin{align}
	J \simeq \dfrac{4}{\sqrt{\pi}} V_s^{3/4} e^{-2\sqrt{V_s}}, \label{Eq:AA-J}
\end{align}
whereas the disorder potential $\Delta$ depends on both $V_s$ and $V_d$, 
\begin{align}
	\Delta \simeq \dfrac{V_d}{2}e^{-\alpha^2/\sqrt{V_s}}. \label{Eq:AA-Delta}
\end{align}
While such a mapping between continuum and tight-binding models is useful in many respects, it clearly misses a qualitative difference between these two limits: the \iLM{} possesses an SPME while the AA model does not. Of course, for $V_s\gg1$, such an SPME induces an exponentially small intermediate phase and is of no physical significance. In reality, we find that $V_s\gtrsim 8$ suffices to reach the AA limit in most practical situations, {which is where most of the localization experiments have indeed been performed so far, justifying the use of the AA model in their analyses~\cite{Schreiber2015,Bloch2016PRL,luschen2016evidence,Bloch2017PRX}. 
What we are predicting in the current work is that an experimental lowering of $V_s$ will lead to significant qualitative (and quantitative) deviations from the AA results.}

From the perspective of a Wegner flow process~\cite{wegner1994flow,Wilson1993PRD,Wilson1994PRD,kehrein2007flow}, the continuum model contains important corrections to the AA model that break its self-duality, thereby producing an SPME. 
The continuum to tight-binding progression in the model defined by Eq.~\eqref{Eq:IncommensuratePotential} can be studied using the Wegner flow method, and we do this in Appendix~\ref{Appendix:WegnerFlow}, establishing the main corrections to Eqs.~\eqref{Eq:AAModel}-\eqref{Eq:AA-Delta} which arise as one flows away from the tight-binding limit.  
This Wegner flow calculation (Appendix~\ref{Appendix:WegnerFlow}) brings out the leading order physics left out in the tight-binding AA limit contributing to the emergence of the SPME in the continuum limit. We find that the Wegner flow indicates the leading order correction to be an on-site term, not a second-nearest neighbor hopping, leading to the deviation from the tight-binding AA limit.
We refer the readers to Appendix~\ref{Appendix:WegnerFlow} for details about our analysis based on the Wegner flow method. 
Since we diagonalize the full continuum Hamiltonian defined by Eq.~\eqref{Eq:IncommensuratePotential} anyway to get our results, the Wegner flow analysis (Appendix~\ref{Appendix:WegnerFlow}) here only provides a theoretical interpolation between the AA model and the full incommensurate model.

The important question is then whether such a qualitative difference between the two models [i.e., tight-binding AA and continuum Eq.~\eqref{Eq:IncommensuratePotential}] can be detected in typical optical lattice potentials available experimentally. 
Our numerical results suggest that this difference is indeed appreciable, as shown in Fig.~\ref{Fig:Imbalance-Vs8}(b), which compares the density imbalance for the \AAM{} and the \iLM{}. 
One can see that while the quantity $\mI$ is rather similar in the two models, the edge density fraction $\mD$ draws a clear distinction between them: the crossover region for $\mD$ in the \iLM{} is noticeably wider. 
Therefore, even in this relatively deep lattice limit we can already expect to observe a departure from the AA model toward the bichromatic incommensurate potential model. 

{
We make a few remarks here. 
First, as noted before, the localization transition in the AA model is rounded out in our numerical results because of both the finite system size and the rational $\alpha$ we use. 
{We do not investigate system size dependencies or the thermodynamic limit here because this work is primarily aimed at making connections to the experiments, which always deal with finite systems. More importantly, the difference between AA model and the continuum model already becomes appreciable in the finite-size systems we study, and thus can already be observed in current experiments without worrying about the thermodynamic limit (since our system size is chosen to be comparable to the typical optical lattice experimental system sizes).}
Second, our results show clearly that it is important to measure both $\mI$ and $\mD$ in order to establish the existence of an SPME (and the corresponding intermediate phase) in the experiment; measuring the density imbalance $\mI$ alone is not able to resolve this issue. 
In fact, our result in Fig.~\ref{Fig:Imbalance-Vs8}(b) shows that the quantity $\mI$ in the \iLM{} behaves almost the same as that in the AA model. 
Therefore, it is crucial that both $\mI$ and $\mD$ are measured in the experiment so as to properly diagnose a possible SPME and the resulting intermediate phase. 
Finding $\mI$ and $\mD$ both to be finite is the definitive evidence for the existence of the intermediate phase associated with the SPME.
{Finally, in many of our numerical results, along with presenting our main results for the continuum incommensurate model of Eq.~\eqref{Eq:IncommensuratePotential}, we also show the results for the corresponding tight-binding AA model as defined by Eq.~\eqref{Eq:AAModel} [using Eqs.~\eqref{Eq:AA-J} and~\eqref{Eq:AA-Delta} for the transformation of the continuum parameters $V_s$, $V_d$ to the AA parameters $\Delta$ and $J$] for the sake of comparison without elaborating on the AA results much in our discussion (since the AA limit has already been studied extensively in the literature going back almost forty years).}
} 

\subsection{The intermediate phase for smaller \texorpdfstring{$V_s$}{Vs}}

\begin{figure}[!]
\includegraphics[scale=1]{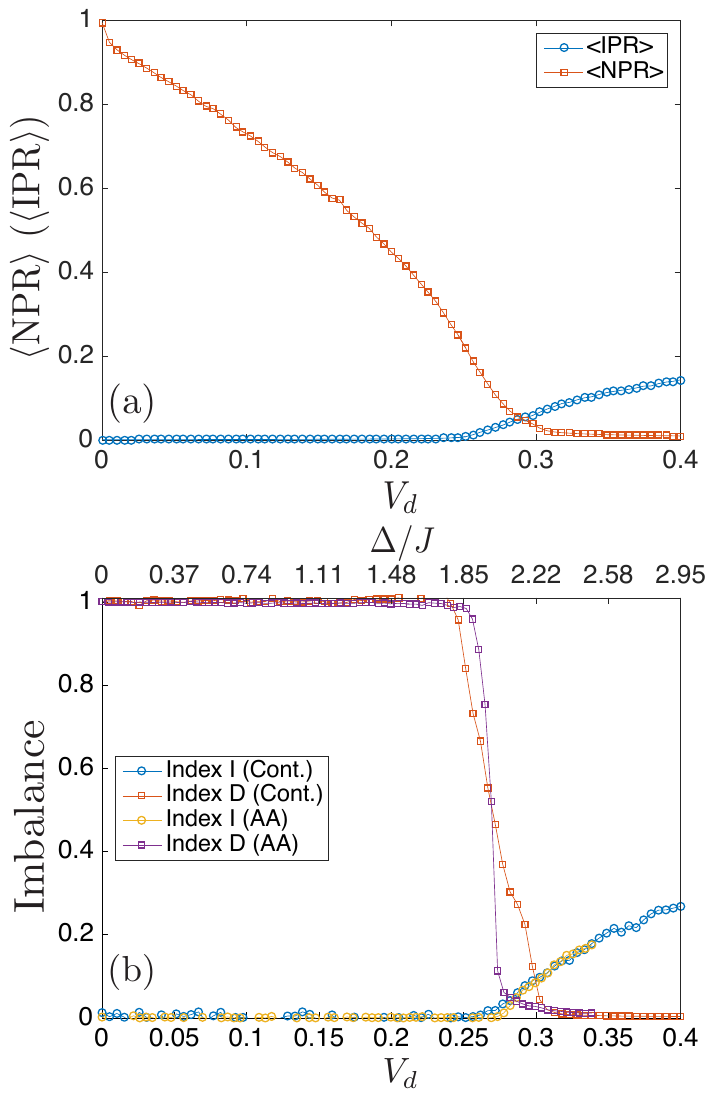}
\caption{\label{Fig:Imbalance-Vs6} 
(a) Averaged NPR and IPR and (b) density imbalance and edge density fraction for the \iLM{} with $V_s = 6$, $\alpha = 532/738$, and a system size of $L=738$.  
The two sets of data in (b) compare the results from the \iLM{} and the AA model, respectively. 
}
\end{figure} 

\begin{figure}[!]
\includegraphics[scale=1]{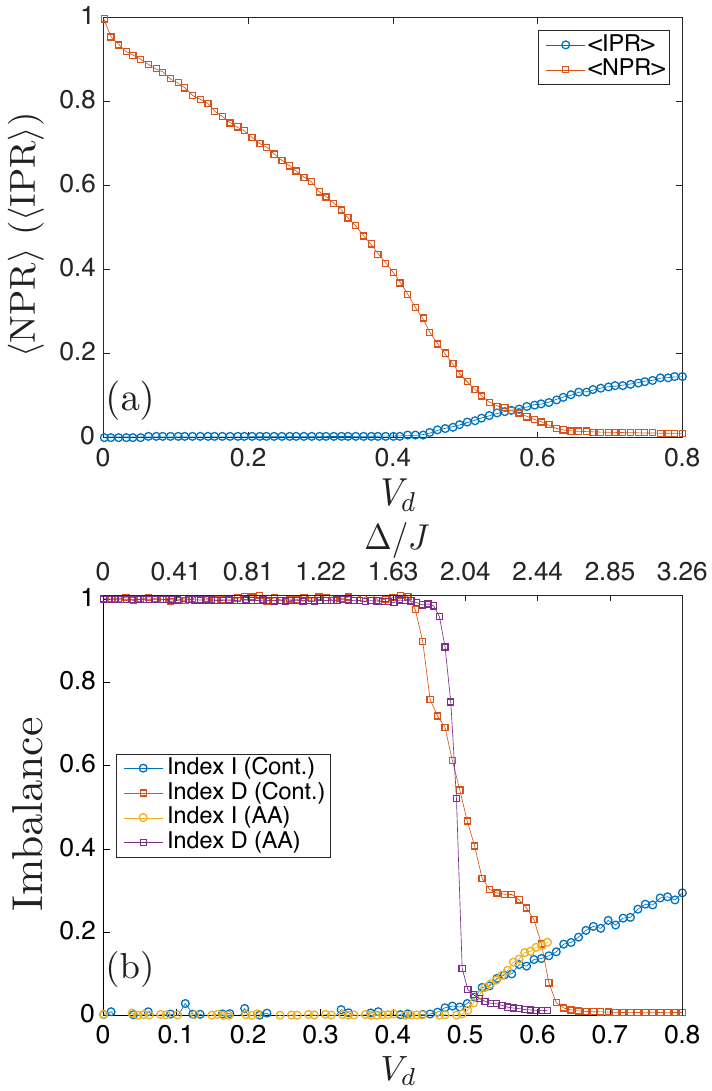}
\caption{\label{Fig:Imbalance-Vs4} 
(a) Averaged NPR and IPR and (b) density imbalance and edge density fraction for the \iLM{} with $V_s = 4$, $\alpha = 532/738$, and a system size of $L=738$.  
The two sets of data in (b) compare the results from the \iLM{} and the AA model, respectively. 
}
\end{figure} 
The essence of the intermediate phase in the \iLM{} is that all eigenstates of the lowest band cannot be localized unless the strength of the ``disorder'' potential $V_d$ overcomes the bandwidth of the lowest band, leading to a possible coexistence of localized and delocalized eigenstates (see also Appendix~\ref{Appendix:SPME}). 
As the bandwidth of the lowest band grows exponentially when $V_s$ is decreased, one naturally expects that the intermediate phase continues to widen with decreasing $V_s$. As a result, it may be advantageous to search for this intermediate phase using a shallower primary lattice. 
This expectation is indeed confirmed by our numerical results. 
For example, Fig.~\ref{Fig:Imbalance-Vs6} shows the density imbalance for $V_s = 6$, which indeed has a wider intermediate phase than that for $V_s = 8$. 
Figure~\ref{Fig:Imbalance-Vs4} further shows the density imbalance for $V_s = 4$, {which has the widest intermediate phase of the three cases we study. We have checked that this trend continues to $V_s =2$ (not shown)}.

\begin{figure}[!]
\includegraphics[scale=1]{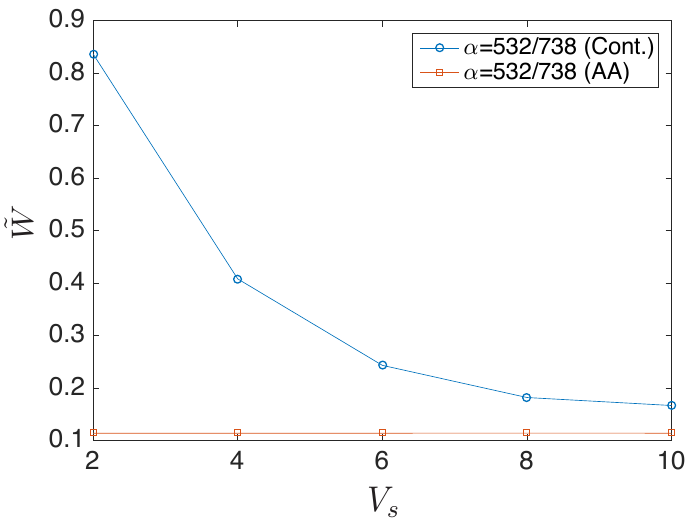}
\caption{\label{Fig:Imbalance-Ratio} 
Relative width $\tilde{W}$ of the intermediate phase as a function of the primary lattice depth $V_s$. Note that the corresponding AA result shows an almost constant dimensionless width of a very small value which provides a measure of the finite size effect. 
}
\end{figure} 

In order to systematically show the width of the intermediate phase as a function of primary lattice depth $V_s$, we introduce a dimensionless relative width $\tilde{W}$ as follows,
\begin{align}
	\tilde{W} = \dfrac{2(V_d^{+}-V_d^{-})}{V_d^{+}+V_d^{-}}, 
\end{align}
where $V_d^{-}$ and $V_d^{+}$ are the lower and upper bound of the intermediate phase, respectively. As shown in Fig.~\ref{Fig:Imbalance-Ratio}, the width of this phase grows significantly with reducing $V_s$. 
Thus, we conclude that using a shallower primary lattice will make this intermediate phase more accessible in the experiment. 
Figure~\ref{Fig:PhaseDiagram_iLM} further summarizes the results in this section in a phase diagram. 
We point out that the results shown in Fig.~\ref{Fig:Imbalance-Ratio} and~\ref{Fig:PhaseDiagram_iLM} indicate that the AA limit probably {becomes more or less reasonable for $V_s>8$}, {but experimental resolution issues in the measurement of $\mI$ and $\mD$ become important in one's ability to determine the intermediate phase for comparatively deeper lattices ($V_s$ $\sim$ $8$ or deeper).} 
Strictly speaking, the AA limit applies only for infinitely deep lattices, and the mobility edge exists for any bichromatic 1D incommensurate potentials represented by Eq.~\eqref{Eq:IncommensuratePotential}, but for all practical purposes the tight-binding AA limit is perhaps reached already for $V_s=8$.

\begin{figure}[!]
\includegraphics[scale=1]{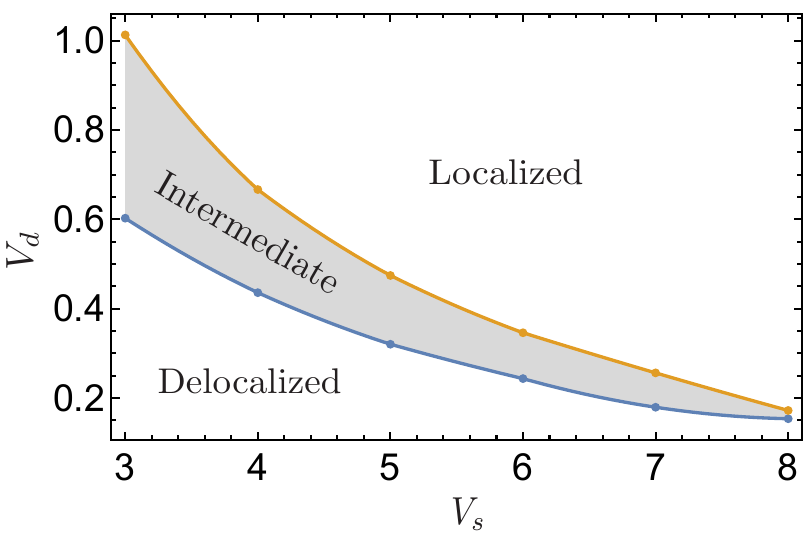}
\caption{\label{Fig:PhaseDiagram_iLM} 
Phase diagram for the \iLM{} with $\alpha = 532/738$ and a system size of $L=738$. The shaded region highlights the intermediate phase due to SPME in this model. The dots are obtained from our numerics, while the solid lines are interpolations which serve as a guide to the eye. 
Note that the intermediate phase essentially vanishes for $V_s>8$ which defines the tight-binding AA limit in this context.
}
\end{figure} 

{
Before we conclude this section, we mention that the presence of an SPME presents additional complications in understanding the MBL transitions of continuous disordered systems. 
Previous theoretical studies  have found non-ergodic metallic phases and full many-body localization for systems of interacting localized and mobile degrees of freedom~\cite{Xiaopeng2015PRL,Mukerjee2015PRL}, although a consensus is still lacking regarding the thermodynamic limit~\cite{Rahul2015PRB,Singh2015PRL,Luitz2015PRB,Roeck2016PRB,Potter2016PRB,Sagi2016PRB,DasSarma2017PRB,khemani2016critical,Roeck2016PRB,Bauer2017PRB}. 
Our finding of a generic SPME in continuous systems makes  understanding the interplay of interactions and SPME unavoidable for characterizing MBL transitions in incommensurate optical lattices. (This is in contrast to the pure Anderson model with random on-site disorder which does not have an SPME.) 
{In Appendix~\ref{Appendix:MBME} we present exact diagonalization results in a small interacting system to show that the intermediate phase can possibly survive finite interactions. However, the eventual fate of the intermediate phase should be tested in the experiment, since our interacting simulations are necessarily in small systems.  
{The system} defined by Eq.~\eqref{Eq:IncommensuratePotential}, which can be implemented experimentally~\cite{Schreiber2015}, is the ideal system to study the interplay of MBL and SPME, which is an important open question in the field.}
} 

\section{The \texorpdfstring{$sp$}{sp} ladder two-band model \label{Section:spModel}}

The \iLM{} we discussed in the previous section is likely the most experimentally accessible setup to observe the SPME in one dimension. 
However, there are conceptually simpler ways to construct models that possess an SPME: one just needs to create a coexistence of localized and delocalized states. 

One such possibility is to consider the lowest two bands (i.e., $s$ and $p$ bands) of the \iLM{}, and let the depth of the disorder lattice potential be just larger than the bandwidth of the lowest band. As a result, all eigenstates in the lowest ($s$) band are completely localized, while all eigenstates in the second ($p$) band are completely delocalized. 
Basically, we have artificially superimposed two tight-binding AA models with the intermediate phase being defined as the situation with one (the $s$-band) being in the localized phase and the other (the $p$-band) being in the delocalized phase. 
We then consider a situation where atoms can be pumped into the second band, say by Raman techniques. 
Then the model Hamiltonian describing such a system is a two-band \iLM{}, $H = H_0 + H_\text{int}$, where 
\allowdisplaybreaks[4]
\begin{align}
	H_0 &= \sum_{j}\left(-t_sc^\dagger_{s,j} c_{s,j+1} + t_p c^{\dagger}_{p,j}c_{p,j+1} + \text{h.c.}\right) \notag\\ 
	&+ \sum_{j}\Delta\cos(2\pi \alpha j + \phi)\left(c^\dagger_{s,j}c_{s,j} + c^\dagger_{p,j} c_{p,j}\right), \notag\\
	H_\text{int} &= V_{sp}\sum_{j}c^\dagger_{s,j}c_{s,j}c^\dagger_{p,j}c_{p,j}. \label{Eq:LadderModel}
\end{align}
This model can also be realized by two species of atoms (say $^{87}$Rb and $^{40}$K) loaded in the incommensurate lattice. Because $^{87}$Rb is a bosonic atom, we shall consider a deep lattice where these atoms behave like hard-core bosons, equivalent to fermions in one dimension. In this model, the particle numbers in the two bands $s$ and $p$, $N_s$ and $N_p$, are separately conserved. In our exact diagonalization calculation, we fix $N_s$ and $N_p$ to be at half filling. 
We also fix $t_p = 2t_s$, which we expect to be a reasonable choice for the K-Rb mixture. 

The physics is now conceptually simple: one band (the $p$-band) is extended and the other band (the $s$-band) localized, and hence the system as a whole is by definition in an intermediate phase.  
By tuning the band occupancies $N_s$ and $N_p$, the system can be continuously tuned from being completely delocalized ($N_s$ zero and $N_p$ finite) to being completely localized ($N_s$ finite and $N_p$ zero) through the intermediate mixed phase (both $N_s$ and $N_p$ finite).  An easier technique is simply to tune the potential strength $\Delta$ so that the system goes between all states delocalized (small $\Delta$) to all states localized (large $\Delta$) with an intermediate state in between.  This is bound to happen whenever the two lattices have different hopping parameters ($t_s \neq t_p$).

To diagnose the ergodicity breaking transition due to the emergence of the intermediate phase, we calculate the evolution of the number imbalance  for this two-band model,
\bea
&& I(\tau)=  \frac{1}{N_s + N_p} \times \nn \\
&& \left\{ \sum_{j \in {\rm even} }[ \langle c_{s, j} ^\dag c_{s, j} \rangle +\langle c_{p, j} ^\dag c_{p, j} \rangle]
			-\sum_{j \in {\rm odd } } [ \langle c_{s, j} ^\dag c_{s, j}  \rangle + \langle c_{p, j} ^\dag c_{p, j} \rangle ] \right\}. 
			\nn 
\eea
To distinguish ergodicity versus non-ergodicity, we choose an initial state with all atoms in the even sites, i.e., $I(\tau = 0 ) = 1$. 
The long-time average of number imbalance vanishes in an ergodic state, but remains finite in a non-ergodic phase. 

To diagnose localization versus delocalization, we investigate the domain wall dynamics of an initial state with all atoms prepared in half 
of the system, say on sites $1\leq i\leq L/2$, with $L$ being the system size. We then monitor the evolution of particle number density at the right end of the lattice, 
\bea
D^{\rm end}  = D_s^{\rm end}  + D_ p  ^{\rm end}, 
\label{eq:D} 
\eea 
where $D_s^{\rm end} = \langle c_{s, L} ^\dag c_{s, L} \rangle$, and $D_p ^{\rm end} =  \langle c_{p, L} ^\dag c_{p, L} \rangle$. 
If $D^{\rm end} $ is finite after long-time average, we have finite diffusion, and the system is delocalized. Otherwise, the system is localized. 

\begin{figure}[!]
\includegraphics[scale=0.8]{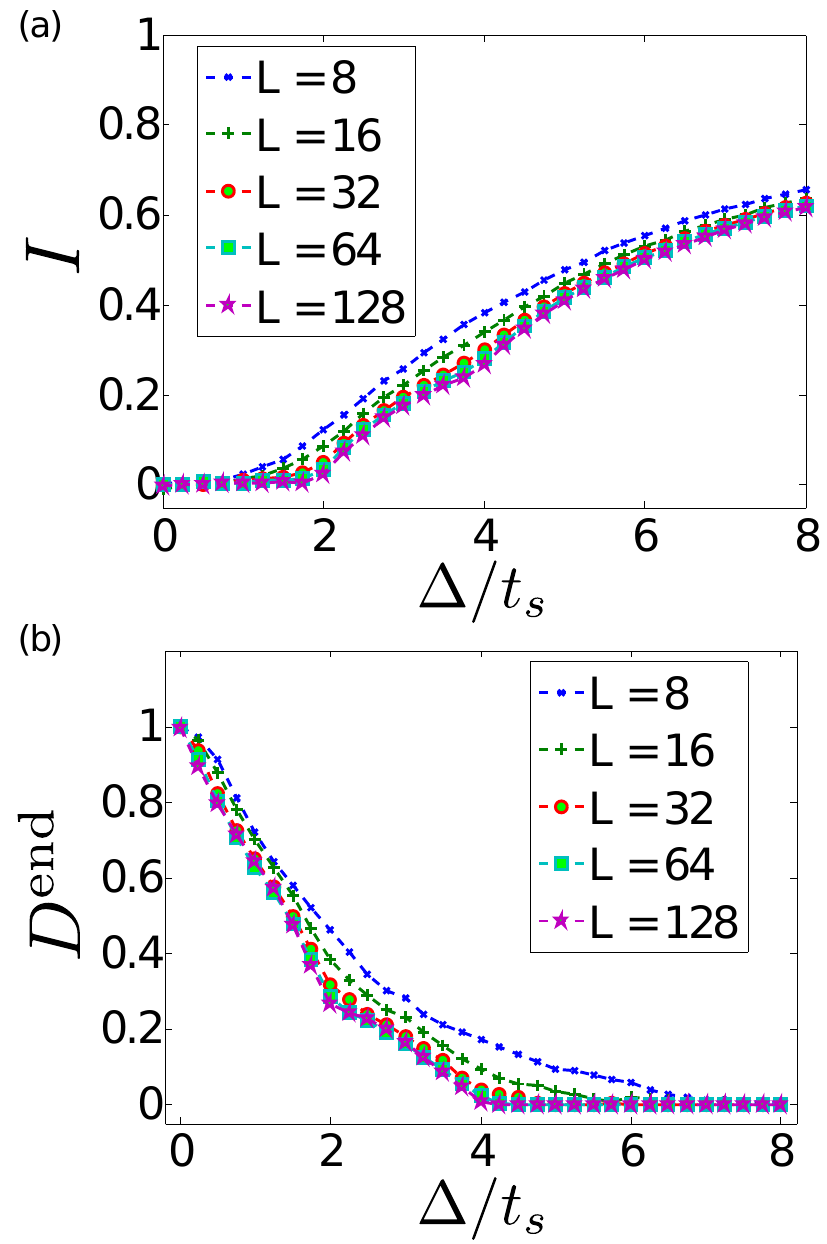}
\caption{\label{Fig:sp-1} 
Localization and ergodic transitions of non-interacting fermions in the two-band \iLM{} at half filling. 
(a) shows the number imbalance ($\mI$) which characterizes the ergodic to non-ergodic transition. $\mI$ becomes finite at the transition from the ergodic to the nonergodic phase. 
(b) shows the edge density $\mD$ (Eq. (4)) characterizing the localization transition. $\mD$ vanishes at the transition from the delocalized to the localized phase. The intermediate region with ($\mathcal{I}\neq0$, $\mathcal{D}\neq 0$) supports a nonergodic extended phase. We use $t_p = 2t_s$ here. The localization transition locates at $\Delta = 4t_s$ and the ergodic transition locates at $\Delta = 2t_s$. 
The region between $2t_s$ and $4t_s$ is the intermediate region by definition as reflected in simultaneous finite values for both $I$ and $D$. 
}
\end{figure} 

It is obvious that $I$ and $D^{\rm end}$ so defined are very similar to the density imbalance $\mI$ and the edge density fraction $\mD$ defined for the bichromatic \iLM{} of earlier sections with the same fundamental meaning and significance---if they are both finite, the system is a nonergodic metal.

\begin{figure}[!]
\includegraphics[scale=0.8]{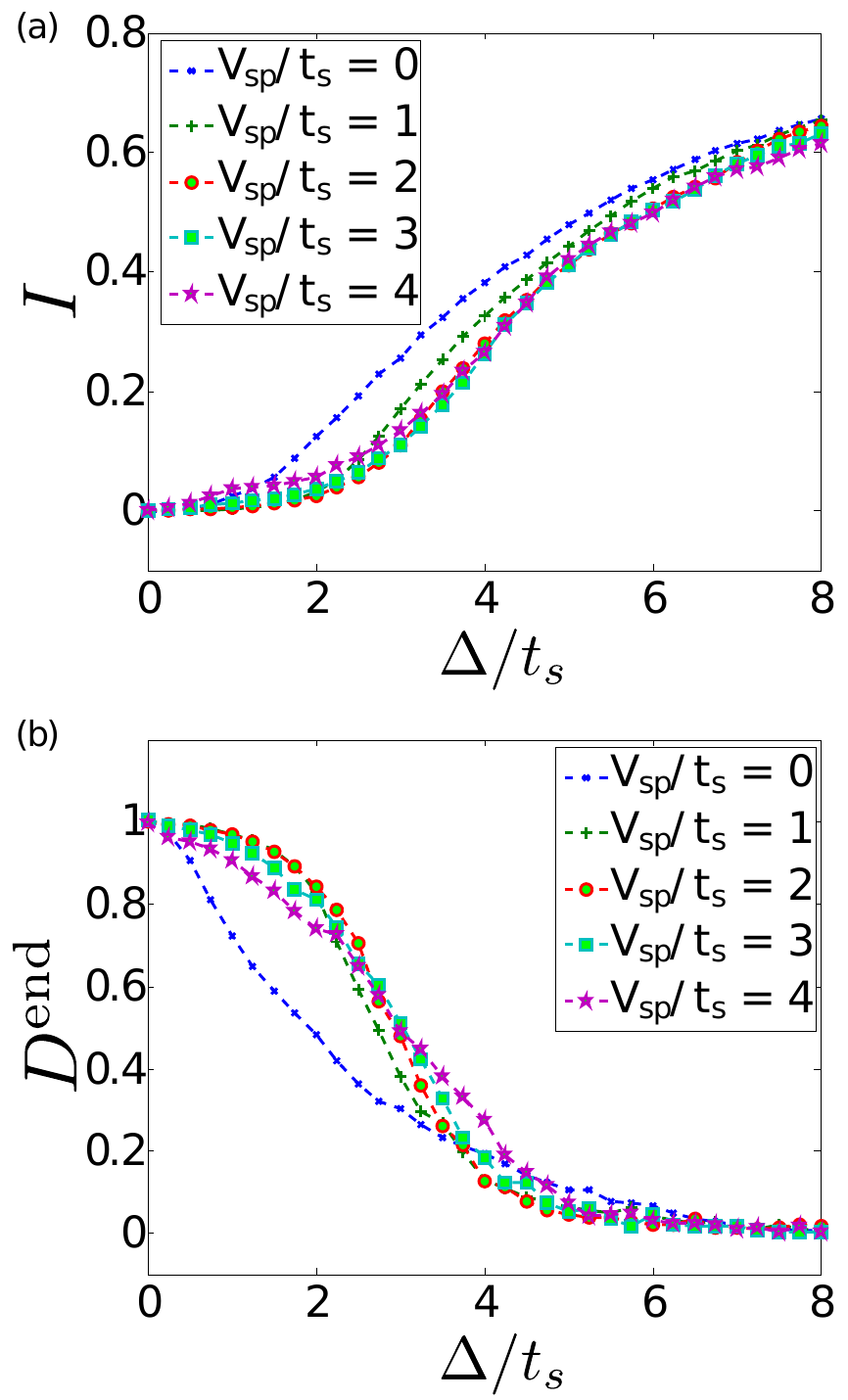}
\caption{\label{Fig:sp-2} 
Localization and ergodic transitions of interacting fermions in the multi-band \iLM{} at half filling. (a) shows the number imbalance $\mI$. (b) shows the edge density $\mD$ [Eq. (4)]. We use $t_p = 2t_s$ and a system size $L = 8$. For the interacting case, our results still suggest there are two distinct transitions. As we increase the incommensurate lattice strength $\Delta$, $\mI$ becomes finite at the ergodic-to-nonergodic transition (here roughly locating at $\Delta/t_s\approx 2$), and the diffusion vanishes (here $\mD$ vanishes) at the delocalization-to-localization transition (here roughly locating at $\Delta/t_s \approx 4$). The parameter region supporting the nonergodic extended phase (with finite number imbalance and finite diffusion, i.e., $\mathcal{I}\neq 0$ and $\mathcal{D}\neq 0$) is considerably large.
}
\end{figure}

\subsection{The noninteracting result}
For non-interacting fermions, we can simulate the dynamics  in systems of large sizes. Figure~\ref{Fig:sp-1} shows the behavior of ${I}$ and $D^{\rm end}$, where we find three distinct regions. 
For small $\Delta$, we have $I =0$ and $D^{\rm end} \neq 0$, which implies the system is ergodic and delocalized. In the opposite limit for large $\Delta$, we have $I \neq 0$ and $D^{\rm end} =0$, indicating that the system is localized and nonergodic. 
It is clear that we also have an intermediate region where both $I$ and $D^{\rm end}$ are finite, which means that the system is non-ergodic but delocalized. For this non-interacting system, the intermediate phase arises because we only have local integrals of motion in the $s$-orbital channel, which is not enough to make the whole system locally integrable. 
Physically, it simply means that the $s$-atoms are localized and cannot move whereas the $p$-atoms are mobile in this intermediate phase, and as we see in Fig.~\ref{Fig:sp-1}, this obvious fact is clearly manifested in the physical quantities $I$ and $D^{\rm end}$.

\subsection{Interacting results in a small system}  
For interacting fermions, we simulate the dynamics with the many-body exact diagonalization method, which is limited to small system sizes due to the exponential scaling of the total Hilbert space dimension. Due to finite-size effects, we cannot reach a sharp conclusion about whether the intermediate phase found for non-interacting fermions in the two-band model is robust against interactions in the thermodynamic limit.  But the numerical results (Fig.~\ref{Fig:sp-2}) indicate that the intermediate phase may survive in the interacting system or at least causes a broad crossover for the MBL transition. 
{We also want to point out that the fact that the ergodic and localization transitions are not strongly modified by interactions (i.e., they still occur in the vincinity of $\Delta/t_s = 2$ and $\Delta/t_s = 4$, respectively) is partially due to the special interaction form in this two-band model [Eq.~\eqref{Eq:LadderModel}]. 
Specifically, there is no interaction between purely localized ($s$-band) and purely extended ($p$-band) particles in our model. In contrast, a more general interacting Hamiltonian [e.g., the one we introduced for the \iLM{} in Eq.~\eqref{Eq:GAAInteractionModel}] may contain such {interaction terms}.
As a result, we expect that the boundary for the ergodic and localization transitions can be modified by interactions. 
If future experiments necessitate a more general interacting Hamiltonian, it is easy to do so within the constraint of small system exact diagonalization studies as carried out in the current work without an $s$-$p$ interaction coupling. 
}

The interplay of MBL and SPME remains an important open question which can be experimentally well-studied in incommensurate lattices as shown explicitly in this section and the last section by studying either the bichromatic \iLM{} in the shallow limit (the last section) or the two-band double-AA model (this section). We emphasize that assuming that Eq.~\eqref{Eq:IncommensuratePotential} represents the AA model with no SPME is strictly speaking incorrect, and all experimental localization studies in incommensurate lattices must address the fact that 1D incommensurate potentials generically allow the existence of SPMEs and intermediate nonergodic metallic phases (albeit with an exponentially suppressed intermediate phase in the deep lattice limit).

\section{Conclusion \label{Section:Summary}}
To summarize, we systematically study the localization physics in the one-dimensional bichromatic incommensurate optical lattice, which has been implemented in cold atom experiments to study single-particle and many-body localization. This system was previously treated using a tight-binding approximation which is valid only in the infinitely deep lattice potential limit, and thus expected to have single-particle degrees of freedom either all localized or all delocalized based on the AA model description. However, we point out that such AA model based analysis is generally incorrect, and that higher-order corrections beyond the AA model approximation are  significant near the localization transition because they generically give rise to mobility edge physics~\cite{PhysRevA.64.033416,PhysRevA.75.063404} not captured by the AA model. 
A Wegner flow analysis clearly brings out how the corrections to the tight-binding limit produce an SPME in the noninteracting system. 
Our theoretical finding provides an important impetus for using incommensurate optical lattices to study the interplay between many-body localization and single-particle mobility edges in 1D systems.  
To fully understand the MBL transition in the experimental incommensurate lattice system, it is inadequate to consider simply the AA model with all single-particle degrees of freedom being localized, since strictly speaking, there are always some extended states in the noninteracting system in the thermodynamic limit for any finite lattice potential (i.e., the AA limit never strictly applies). Addressing the question of how localized and extended degrees of freedom mix together in the presence of interactions, as originally raised in Refs.~\cite{Xiaopeng2015PRL,Xiaopeng2016PRB}, is unavoidable in such continuous incommensurate optical lattice systems. This also indicates MBL transitions in lattice models and continuous systems could be drastically different as a matter of principle, particularly in the thermodynamic limit. 
We note the crucial point that the existence of a mobility edge implies that a finite fraction of the full spectrum remains localized (with the other fraction delocalized) even in the thermodynamic limit, i.e., the numbers of localized and delocalized states both diverge in the thermodynamic limit although they arise from different single-particle energies. 

In numerical calculations, we first show the existence of an SPME by calculating the averaged NPR and IPR of all eigenstates in the lowest band. 
These two quantities can be combined to clearly distinguish the three regimes of the model in theory: when the averaged IPR vanishes, the system is completely delocalized; when the averaged NPR vanishes, the system is completely localized; only when both quantities remain finite will the system be in an intermediate phase.  
To connect to experiments, we identify two physical observables in quench dynamics, i.e., the density imbalance $\mI$ and edge density $\mD$. The former is to diagnose the ergodic to non-ergodic transition, as the density operator in general is expected to have overlap with local integrals of motion if the system is nonergodic. The latter is directly related to whether the system is localized or not. Combining them together, we can identify three distinct phases in an unambiguous way with the intermediate nonergodic metallic phase being the phase where both $\mI$ and $\mD$ are non-zero.  

{
Our study also sheds light on the experimental efforts to search for SPME in one-dimensional systems. 
It is well known that non-interacting fermions in a disordered lattice become localized due to quantum interference effects~\cite{Anderson1958} independent of the disorder strength. In one and two dimensions all eigenstates are localized by an infinitesimal amount of random disorder~\cite{Abrahams1979,Lee1985RMP}. 
However, in three dimensions electron localization in disordered systems is characterized by the appearance of an SPME. In fact, the existence of an SPME is a general property in generic three-dimensional disordered systems as well as various one-dimensional \iLM{}s. However, despite its ubiquity in theoretical constructions, the SPME still remains elusive in the experiments, especially in one-dimensional systems. 
We believe that the simple bichromatic \iLM{} (as well as the two-band model) we studied in this paper provides the simplest suitable experimental candidate for the clear observation of a one-dimensional SPME, because it only requires slight modifications of existing experimental setups. 
All one needs to do is to measure $\mI$ and $\mD$ simultaneously as a function of $V_d$ to see the clear signature of SPME.  Turning the interaction on in such a system then will tell us about the interplay between interaction and mobility edge in the context of many body localization phenomena.}

We conclude by emphasizing our four new results in the paper: (1) we show that 1D bichromatic incommensurate potentials manifest single-particle-mobility-edges and intermediate phases (which are neither completely localized nor completely extended) in the noninteracting limit, also establishing the multifractal critical properties of the SPME and the intermediate phase; 
(2) we show that this mobility edge and the associated intermediate phase remain stable to interactions (in small systems) manifesting a many body localization transition; 
(3) we propose realistic cold atom optical lattice experiments in two distinct systems (bichromatic incommensurate potential and two-band $s$-$p$ ladder) to observe the predicted single-particle-mobility edge as well as to study the many-body-localization properties for the corresponding interacting system; 
(4) an obvious important consequence of our theory is that one-dimensional bichromatic incommensurate lattice systems can be experimentally (and numerically) used to directly study the interplay between single-particle mobility edge and many body localization in realistic optical lattices-- in particular, the open question of whether a system with a single particle mobility edge can or cannot manifest many body localization becomes experimentally accessible along with clear protocols for a practical search for the nonergodic extended phase (i.e., the intermediate phase) in interacting disordered many-body systems.

\paragraph*{Acknowledgment.----}  
The authors thank Henrik L\"uschen and Immanuel Bloch for discussions and helpful comments on the manuscript.
Xiao Li would also like to thank Yizhuang You for helpful discussions. 
This work is supported by JQI-NSF-PFC and LPS-MPO-CMTC. 
In addition, Xiao Li acknowledges the PFC seed grant at JQI for support. 
Xiaopeng Li acknowledges the support by the Start-up Fund of Fudan University. 
The authors acknowledge the University of Maryland supercomputing resources made available for conducting the research reported in this paper.

\appendix

{
\section{The effect of a finite harmonic trap \label{Section:Trap}}

\begin{figure}[!]
\includegraphics[scale=0.7]{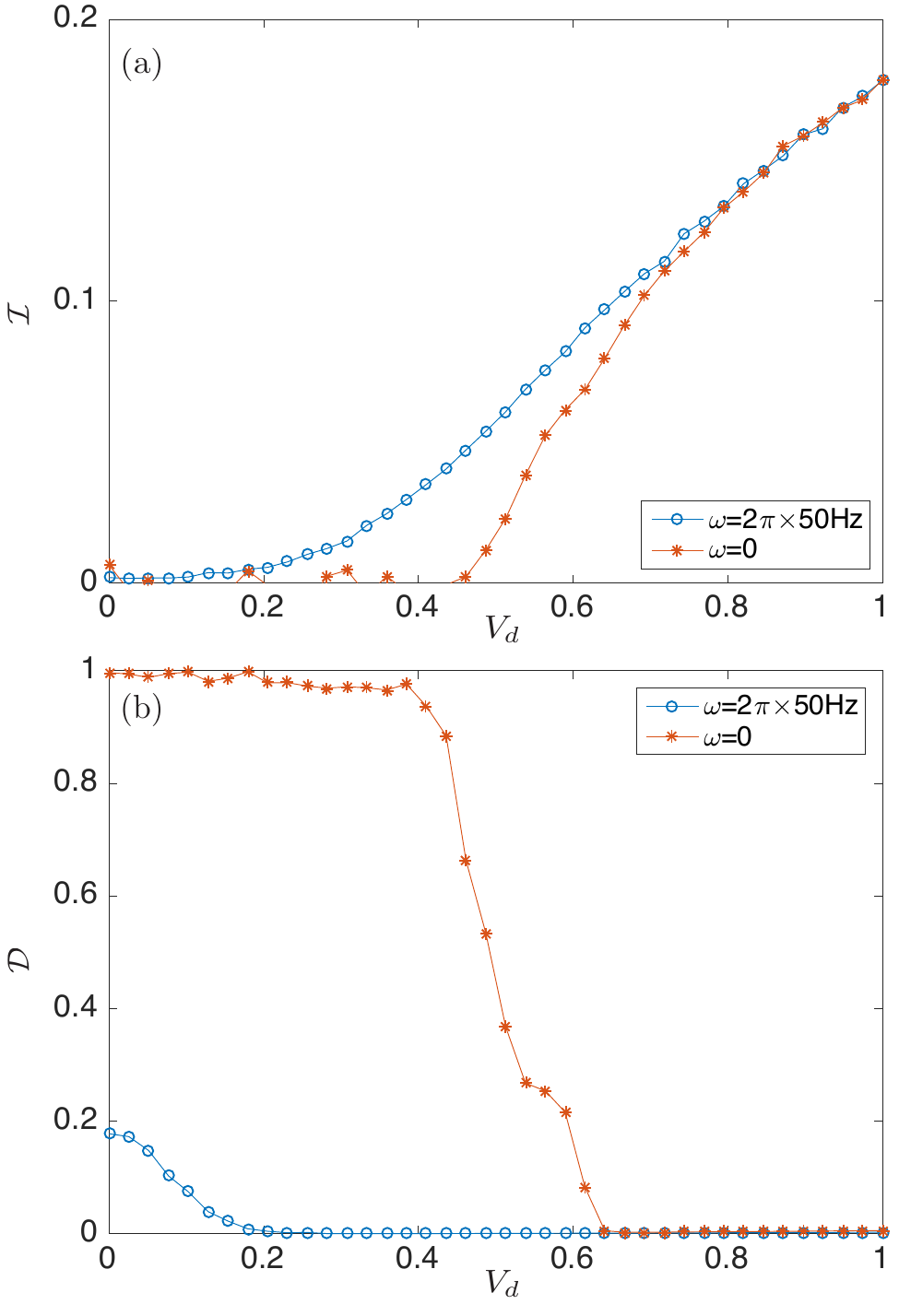}
\caption{\label{Fig:Imbalance-Appendix} Effects of a finite harmonic trap on the density imbalance measurements. The strength of the harmonic trap potential is $\omega = 2\pi\times \SI{50}{Hz}$. 
In addition, the system size is $L=369$, $\alpha=532/738$, and $V_s=4$.}
\end{figure}

In the usual cold-atom experiments the atoms are held in a harmonic trap potential. 
Such a trap potential naturally produces a finite localization length for the atoms, and thus it is natural to ask how it affects the experimental scheme we proposed here to observe the SPME. 
It turns out that leaving the full trap potential on is detrimental for our proposal. 
To show this, we add a harmonic trap potential to the continuum Hamiltonian in Eq.~\eqref{Eq:IncommensuratePotential}, and evaluate the density imbalance $\mathcal{I}$ as well as the edge density fraction $\mathcal{D}$. 
We choose the strength of the trap to be $\omega = 2\pi\times\SI{50}{Hz}$, which is the value realized in recent experiments~\cite{Schreiber2015}. 
As shown in Fig.~\ref{Fig:Imbalance-Appendix}(a), the harmonic trap potential produces a nonzero imbalance even when the pristine system (i.e., the one without the trap potential) is still delocalized. Such an effect can be explained by the fact that the harmonic trap produces a large energy barrier for the particles, especially those far away from the trap center, leading to some degree of localization even in the absence of the disorder potential. 
The harmonic trap has an even stronger effect on the measurement of the edge density fraction $\mathcal{D}$. As shown in Fig.~\ref{Fig:Imbalance-Appendix}(b), the harmonic trap severely hinders the expansion of the initial cloud, as the edge density quickly drops to zero as $V_d$ increases, despite the fact that the pristine system is still delocalized. 

The above results show that a relatively flat potential landscape is needed to achieve the experimental scheme we propose in this work. 
However, such a requirement is well within the current experimental capabilities~\cite{FlatTrap1,FlatTrap2}. Thus, we believe that our proposal remains very feasible experimentally. 
}

\section{Connection to the \AAM{} model from a Wegner flow perspective \label{Appendix:WegnerFlow}}

In this appendix we use the Wegner flow method to systematically generate higher-order corrections to the AA model, which will serve as an effective tight-binding Hamiltonian for the bichromatic \iLM{} in Eq.~\eqref{Eq:IncommensuratePotential}. 
As we show in the following, the correction responsible for the single-particle mobility edge only appears at the second order in the perturbation theory. 

\subsection{The Wegner flow method}
The Wegner flow method was initially developed~\cite{wegner1994flow,Wilson1993PRD,Wilson1994PRD,kehrein2007flow} for matrix diagonalization, and is related to the double-bracket flow algorithm in mathematics~\cite{Casas2004}. 
One advantage of this method is that it can lower the complexity of the original diagonalization problem, allowing one to study large systems more efficiently. 

For our purpose, it is convenient to carry out the Wegner flow in the momentum space. We start by decomposing the \iLM{} in Eq.~\eqref{Eq:IncommensuratePotential} as $\mH = \mH_0 + \mH_1$, where $\mH_0$ defines the pristine periodic lattice, while $\mH_1$ includes the disorder potential proportional to $V_d$. In the momentum space $\mH_0$ has a simple form in terms of Bloch modes, 
\begin{align}
	\mH_0 = \sum_{n,\bk} \varepsilon_n(\bk)\phi_n^\dagger(\bk)\phi_{n}(\bk), 
\end{align}
where $n$ is the band index, $\bk$ is the lattice momentum, and $\phi_{n}(\bk)$ is the Bloch wave function. 
The incommensurate lattice potential can be generally written in the same basis as follows, 
\begin{align}
	\mH_1 = \sum_{nn',\bk\bk'} \mathcal{W}_{nn'}(\bk,\bk')\phi^\dagger_{n}(\bk)\phi_{n'}(\bk'), 
\end{align}
which will induce both interband and intraband transitions. To restore the band concept in the presence of the disorder potential, we {would like to} recast the entire Hamiltonian $\mH$ into a block-diagonal form, with each block taking the form 
\begin{align}
	\tilde{\mH} = \sum_{n,\bk,\bk'}\mathcal{E}_{n}(\bk,\bk')\tilde{\phi}_{n}^{\dagger}(\bk)\tilde{\phi}_{n}(\bk'). \label{Eq:FinalForm}
\end{align}
This can be achieved via a Wegner flow process, which implements a unitary transformation on the Hamiltonian $\mH$ (and thus keeping its spectrum invariant). Specifically, the procedure starts by introducing a matrix $\mathcal{K}$, which is governed by the following equation, 
\begin{align}
	\dfrac{d\mathcal{K}}{dl} = \left[[\mathcal{K}(l),\mathcal{N}], \mathcal{K}\right], \label{Eq:FlowEquation}
\end{align}	
where $\mathcal{N}$ is a diagonal matrix, 
\begin{align}
	\mathcal{N}_{nn'}(\bk,\bk') = \bar{\varepsilon}_{n}\delta_{nn'}\delta_{\bk\bk'}, 
\end{align}
and $\bar{\varepsilon}_{n} = L^{-1}\sum_{\bk}\varepsilon_n(\bk)$ is the average energy of all eigenstates in the $n$th band. The initial condition of the flow is given by 
\begin{align}
	\mathcal{K}_{nn'}(\bk,\bk')\rvert_{l=0} = \varepsilon_n(\bk)\delta_{nn'}\delta(\bk\bk') + \mathcal{W}_{nn'}(\bk,\bk'). 
\end{align}
This flow equation then generates a unitary transformation $\mathcal{K}(l) = O(l)\mathcal{K}(0)O^\dagger(l)$, with 
\begin{align}
	O(l) = T_{l}\exp\left(\int_0^l dl [\mathcal{N},\mathcal{K}(l)]\right), 
\end{align}
where $T_l$ is the time ordering operator. 
Such a process will continuously bring the Hamiltonian $\mH$ into a block-diagonal form. If we denote the final result of the flow as $\tilde{\mathcal{K}}\equiv \mathcal{K}(\infty)$, the block-diagonal matrices in Eq.~\eqref{Eq:FinalForm} are given by 
\begin{align}
	\mathcal{E}_{n}(\bk,\bk') = \tilde{\mathcal{K}}_{nn}(\bk,\bk'). 
\end{align}
Such a procedure also provides a unitary transformation that relates $\phi_{n}(\bk)$ and $\tilde{\phi}_{n}(\bk)$ as $\tilde{\phi}_{n}(\bk) = U\phi_{n}(\bk)$, with $U = O(\infty)$. 

\subsection{Effective lattice Hamiltonian for the \iLM{}}
We now use the Wegner flow procedure to derive an effective Hamiltonian for the \iLM{} in the tight-binding limit, where $V_d\ll V_s$. We will show that the higher-order corrections to the AA model are vital to capture the mobility edge inherent in the \iLM{}. 
{ For the convenience of our discussions below, we redefine the AA model parameters as follows, 
\begin{align}
	Eu_n = -t(u_{n-1}+u_{n+1}) + \mathcal{V}_1\cos(2\pi\alpha n+\phi) u_n. 
\end{align} 
We will see that corrections to the hopping ($t$) and on-site energy ($\mathcal{V}_1$) are generated order by order in the Wegner flow. 
}

We start by writing the matrix $\mathcal{K}(l)$ as a series, 
\begin{align}
	\mathcal{K}(l) = \mathcal{K}^{(0)} + g\mathcal{K}^{(1)} + g^2\mathcal{K}^{(2)} + \dots, 
\end{align}
where $g$ keeps track of the order of perturbative expansion. 
Meanwhile, in this tight-binding limit we can write the incommensurate lattice Hamiltonian formally as $\mH = \mH_0 + g\mH_1$. We can then solve the flow equation in Eq.~\eqref{Eq:FlowEquation} perturbatively and obtain $\mathcal{K}$ [and thus $\mathcal{E}_n(\bk,\bk')$] to all orders. In particular, we find that up to second order in $g$ the block-diagonal matrix $\mathcal{E}_n(\bk,\bk')$ has the following form, 
\begin{align}
	\mathcal{E}_n(\bk,\bk') &= \varepsilon_n(\bk)\delta_{\bk\bk'} + g \mathcal{W}_{nn}(\bk,\bk') \\
	&+ 2g^2\sum_{m\neq n,\bq}\dfrac{\mathcal{W}_{nm}(\bk,\bq)\mathcal{W}_{mn}(\bq,\bk')}{\varepsilon_n(\bk) + \varepsilon_{n}(\bk')-2\varepsilon_{m}(\bq)} +\mathcal{O}(g^3). \notag
\end{align}

Such a procedure can produce the following effective lattice Hamiltonian for the \iLM{} in Eq.~\eqref{Eq:IncommensuratePotential}, 
\begin{align}
	H_\text{AA} &= -J_0 \sum_{j, \sigma} (c^\dagger_{j+1,\sigma}c_{j,\sigma}+\text{h.c.}) + \Delta \sum_{j, \sigma}\cos(2\pi\alpha j + \phi) n_{j,\sigma}, \notag\\ 
	H' &= J_1 \sum_{j,\sigma} \cos\left[2\pi\alpha\left(j+\dfrac{1}{2}\right) + \phi\right](c^\dagger_{j+1,\sigma}c_{j,\sigma}+\text{h.c.}) \label{Eq:GAAModel}\\ 
	&+ J_2 \sum_{j,\sigma} (c^\dagger_{j+2,\sigma}c_{j,\sigma}+\text{h.c.}) + \Delta'\sum_{j, \sigma}\cos(4\pi\alpha j + 2\phi) n_{j,\sigma}, \notag
\end{align}
In the above Hamiltonian the operators $c_{j,\sigma}^\dagger$ and $c_{j,\sigma}$ are the creation and annihilation operators for spin $\sigma = \uparrow, \downarrow$ on lattice site $j$. 
The first term $H_\text{AA}$ is the Aubry-Andre Hamiltonian, while the second term $H'$ contains corrections up to second order in the Wegner flow process, which become important in the shallow (primary) lattice limit. 
We can see that these corrections can modify both the on-site disorder potential and the nearest-neighbor hopping term. In addition, second-nearest neighbor hopping terms can also be generated. 

In fact, this lattice Hamiltonian is very useful in capturing the SPME physics in the \iLM{} in Eq.~\eqref{Eq:IncommensuratePotential}. In Fig.~\ref{Fig:LatticePR} we plot the averaged IPR and NPR for this lattice model with $V_s = 4$ and $\alpha = 532/738$. For a given $V_d$ we use Wegner flow to determine the model parameters in the correction term $H'$. 
We find that the intermediate phase derived from this lattice model agrees with the predictions of the \iLM{} in Fig.~\ref{Fig:Imbalance-Vs4}. 
Such an effective lattice Hamiltonian will be useful when we study manybody effects in this \iLM{}, as it is difficult to incorporate electron-electron interactions directly in the continuum model. 

\begin{figure}[!]
\includegraphics[scale=1]{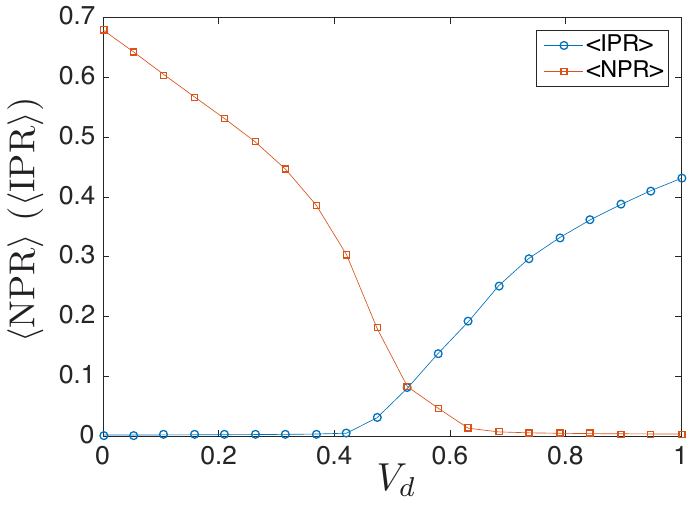}
\caption{\label{Fig:LatticePR} Phase diagram for the generalized AA model in Eq.~\eqref{Eq:GAAModel}. Here we choose $L=738$, $V_s = 4$, and $\alpha = 532/738$. For a given $V_d$, all coefficients in the lattice Hamiltonian [Eq.~\eqref{Eq:GAAModel}] are generated by the Wegner flow process. }
\end{figure}

\begin{figure}[!]
\includegraphics[scale=0.8]{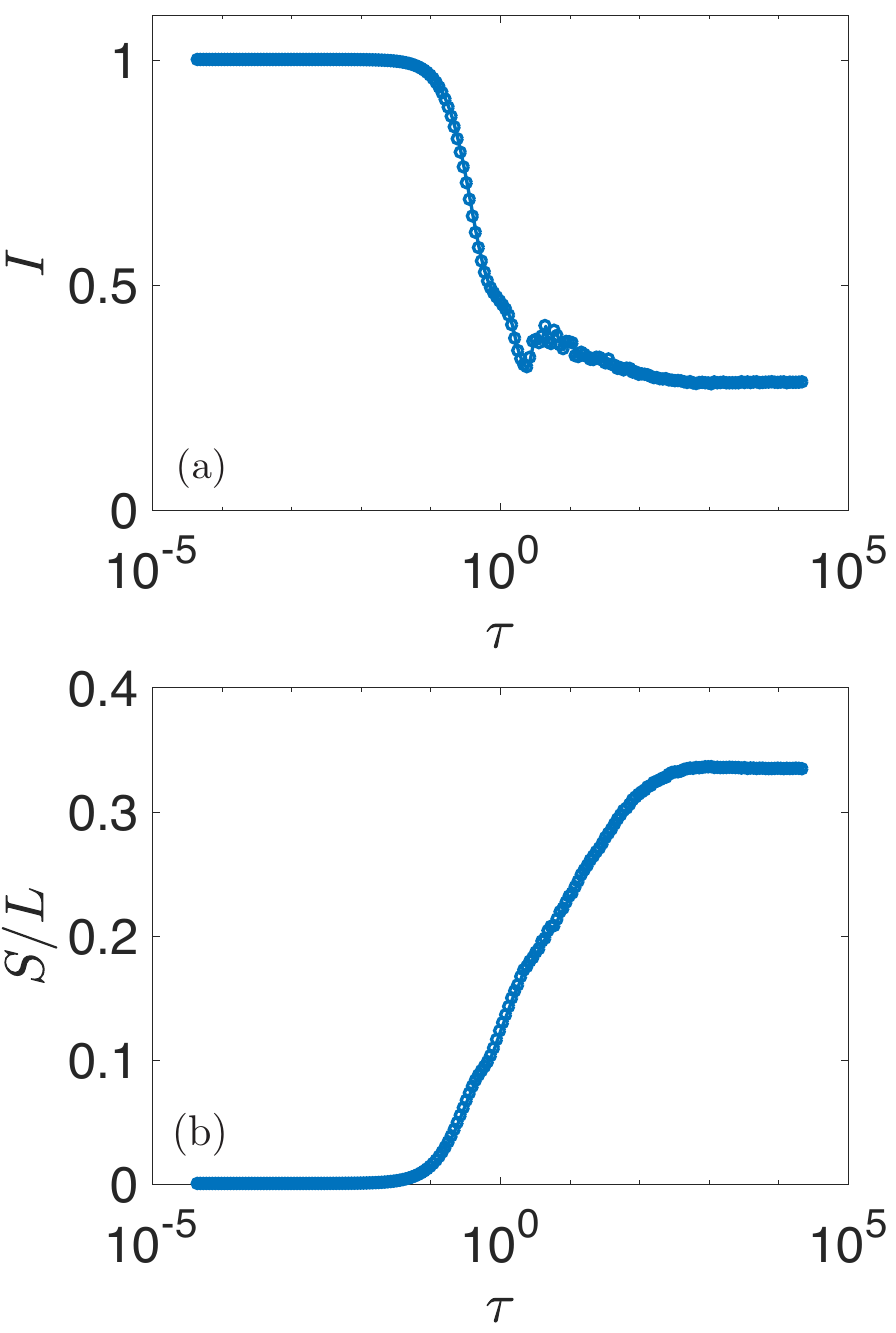}
\caption{\label{Fig:MBME} Evidences for the survival of the intermediate phase in the presence of electron-electron interactions in a small system with $L=8$. (a) shows that the imbalance remains finite in the presence of interactions, while (b) shows the logarithmic growth of entanglement entropy with time. In this calculation we use the model in Eq.~\eqref{Eq:GAAInteractionModel} with $\mathcal{V}_1 = 2t$, $\mathcal{V}_2 = 0.2t$, and $U=5t$. The time $\tau$ is in units of $\hbar/t$. }
\end{figure}

\section{The possibility of a many-body mobility edge \label{Appendix:MBME}}

In this section we provide exact diagonalization results in a small system to show that the intermediate phase we established in the noninteracting 1D bichromatic \iLM{} [Eq.~\eqref{Eq:IncommensuratePotential} in the main text] can possibly survive finite electron-electron interactions. 
The effective tight-binding model we use reads as follows, 
\begin{align}
	H_0 &= \sum_{j,\sigma=\uparrow,\downarrow}\left(-tc^\dagger_{j,\sigma} c_{j+1,\sigma} + \text{h.c.}\right) \notag\\ 
	&+ \sum_{j,\sigma=\uparrow,\downarrow}\left[\mathcal{V}_1\cos(2\pi \alpha j)+\mathcal{V}_2\cos(4\pi \alpha j)\right]c^\dagger_{j,\sigma}c_{j,\sigma}, \notag\\
	H_\text{int} &= U\sum_{j}n_j(n_j-1), \label{Eq:GAAInteractionModel} 
\end{align}
where $n_j = c^\dagger_{j,\uparrow}c_{j,\uparrow}+c^{\dagger}_{j,\downarrow}c_{j,\downarrow}$ is the total electron density on each lattice site, and $\uparrow,\downarrow$ denotes spin-up and spin-down, respectively. 
The above model contains the correction to the AA model we found in Eq.~\eqref{Eq:GAAModel} that is responsible for the intermediate phase. 
We find (see Fig.~\ref{Fig:MBME}) that indeed the intermediate state survives weak interaction in small systems. We show this by calculating both density imbalance and entanglement entropy.

In Fig.~\ref{Fig:MBME} we show the results for a small system with $L=8$ lattice sites. We place the system in the intermediate phase using $\mathcal{V}_1 = 2$ and $\mathcal{V}_2=0.2$, and turn on an interaction strength of $U = 5t$. 
Figure~\ref{Fig:MBME}(a) shows the time evolution of density imbalance $I$, which demonstrates that the initial charge density wave pattern does not relax completely even in the long time limit. 
{Figure~\ref{Fig:MBME}(b) further shows a logarithmic growth of entanglement entropy with time, which is an indication that the system is nonthermal.} 
As a result, we can conclude that at least in this small system the intermediate phase found in the noninteracting model can survive finite electron-electron interactions, giving rise to a possible manybody mobility edge. 
{Moreover, we have carried out an approximate finite size scaling analysis using exact diagonalization up to system sizes of $L=16$, finding that our conclusion about the existence of MBL and a manybody mobility edge remains unchanged.}
The question of what happens in the thermodynamic limit, however, cannot be addressed in our work and remains open for the interacting system.

\bibliography{Bib-MBLattice}
\end{document}